\documentclass[conference, 10pt]{IEEEtran}
\pdfoutput=1
\IEEEoverridecommandlockouts
\usepackage[usenames,dvipsnames]{xcolor}
\usepackage{cite}
\usepackage{epsf}
\usepackage{psfrag}
\usepackage{bbm}
\usepackage{amsmath,amssymb,amsfonts}
\usepackage{algorithmic}
\usepackage{graphicx}
\usepackage{textcomp}
\usepackage{color}
\usepackage{xcolor}

\def\BibTeX{{\rm B\kern-.05em{\sc i\kern-.025em b}\kern-.08em
		T\kern-.1667em\lower.7ex\hbox{E}\kern-.125emX}}
\newtheorem{theorem}{Theorem}

\newtheorem{lemma}{Lemma}
\newtheorem{definition}{Definition}
\newtheorem{remark}{Remark}
\begin{document}
	
	\title{Adaptive Interference Coordination over Channels with Unknown State at the Encoder and the Decoder\\
		{\footnotesize \textsuperscript{}}}
	
	\author{\IEEEauthorblockN{Michail Mylonakis}
		\IEEEauthorblockA{\textit{Division of Inf. Science \& Eng.} \\
			\textit{KTH Royal Institute of Technology}\\
			mmyl@kth.se}
		\and
		\IEEEauthorblockN{ Photios A. Stavrou}
		\IEEEauthorblockA{\textit{Division of Inf. Science \& Eng. } \\
			\textit{KTH Royal Institute of Technology}\\
			fstavrou@kth.se}
		\and
		\IEEEauthorblockN{ Mikael Skoglund}
		\IEEEauthorblockA{\textit{Division of Inf. Science \& Eng.} \\
			\textit{KTH Royal Institute of Technology}\\
			skoglund@kth.se}}

	\maketitle
	
	\begin{abstract}
	We generalize the problem of controlling the interference created to an external observer while communicating over a discrete memoryless channel (DMC) which was studied in \cite{serrano:2014}. In particular, we consider the scenario where the transmission is established over a compound DMC channel with unknown state at both the encoder and the decoder. Depending on the exact state $s$ of the channel, we ask for a different level of average precision $\Delta_s$ on the establishment of the interference coordination with the external observer. For this set-up, we fully characterize the capacity region. 
\end{abstract}
\section{Introduction}
\par Communication in most practical scenarios is subject to undesirable and often unavoidable interference (or disturbance) that relegates the performance of neighbouring transceivers and impairs the operation of nearby electronic devices. A novel communication set-up under disturbance constraints was studied in \cite{bandemer:2011}. Among other results, the authors therein characterize the rate disturbance region for the single disturbance constraint case and bounds for the the two disturbance constraints.
In \cite{serrano:2014}, the authors modeled the communication-induced disturbances in terms of their empirical distribution and investigated the fundamental limits on the communication rate imposed by the requirement that the disturbance in their system model is close enough in probability to a desired type (i.e., empirical distribution). They fully characterized an achievable communication-interference region for the single-user scenario and they derived an achievable region for the multiple-user scenario. For the specific problem, they adopted the general terminology {\it communication and interference coordination}. An interesting observation  regarding \cite{serrano:2014}, is that the choice of interference coordination criterion adopted in therein utilizes ideas from the framework of empirical coordination which was introduced in \cite{cuff:2010}. 
Recently in \cite{cervia:2020}, the authors introduced a stronger version of interference coordination where the requirement is to achieve a certain distribution instead of a type.
The main contribution therein is the complete characterization of the rate region. 
\begin{figure}
	\begin{center}
		
		\includegraphics[width=8cm,height=9cm,keepaspectratio]{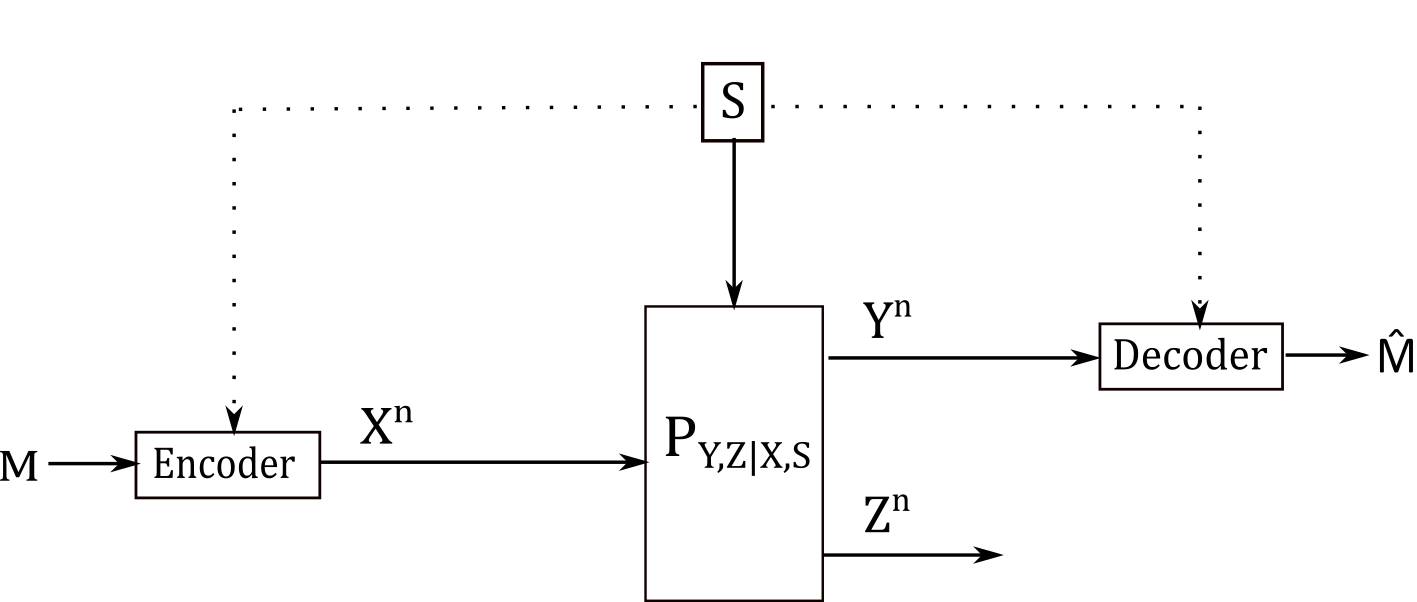}
	\end{center}
	\caption{System Model.}
	\label{fig:fig1}
\end{figure}

\par In this work, we consider the set-up illustrated in Fig. \ref{fig:fig1}. This corresponds to a compound DMC with one input $X$ and two outputs $Y$ and $Z$. The output $Y$ is the observation at the intended receiver, while $Z$ corresponds to an undesired interference created to an external observer. The compound channel is a DMC with a state $S$ which expresses the fact that the channel is selected arbitrarily from a set of possible DMCs and is fixed throughout the transmission block. It models communication in the presence of uncertainty about channel statistics. 
The compound channel was independently studied in \cite{blackwell:1959,dobrushin:1959,wolfowitz:1960}. Generalizations of the capacity formula of the compound channel using the information stable approach can be found in \cite{loyka-charalambous:2016} (see also the references therein).
In this work, we assume that the exact state of the channel is not available at either the encoder or the decoder (dotted lines in Fig. \ref{fig:fig1}). The encoder-decoder pair can use the channel for communicating reliably a random message $M$ as long as the interference is close enough in terms of its type $P_{z^n}$ to a type $Q$. In contrast to the choice of the interference coordination criterion of \cite{serrano:2014}, in our study we make use of the framework of {\it imperfect} empirical coordination. The framework of {\it imperfect} empirical coordination was introduced in \cite{mylonakis:2019} and compared to perfect empirical coordination of \cite{cuff:2010} which asks the joint type of the actions to be close to the desired empirical distribution in probability, it requires the average distance (as measured by total variation) between the joint type of the actions and the desired empirical distribution to be smaller than a certain level $\Delta$ for long enough $n$-length sequences. 
Specifically, in parallel with the establishment of reliable communication between the encoder and the decoder, we ask for a different average level of precision $\Delta_s$ on the establishment of the interference coordination with the external observer, depending on the exact state $s$ of the channel. The specific setup can be applied in the design of more {\it adaptive} communication systems that, for example, require accuracy of different levels for the interference coordination with the external observer, depending on the conditions of the channel, i.e., if the channel is very noisy or less noisy.  
\par  We break the derivation in three steps:
\begin{enumerate}
	\item[(R1)] The characterization of the capacity region for the preliminary problem of {\it multiple} interference coordination over the compound channel (see Theorem \ref{th:compoundperfect});
	\item[(R2)] A theorem which connects the capacity regions for the problems of {\it multiple} and {\it adaptive} interference coordination (see Theorem \ref{th:compoundperfectimperfect});
	\item[(R3)] The characterization of the capacity region for the problem of {\it adaptive} interference coordination over the compound channel (see Theorem \ref{th:compoundimperfect}).
\end{enumerate}

\section{Preliminaries}
\label{s:preliminaries}
We begin with some basic mathematical concepts and the definitions of the matching PMF of a type, the $\Delta$-neighborhood and the  $\Delta$-pre-image, three concepts which will help us in the statement of our results. Furthermore, we define the communication protocol which is used in the set-up. We discuss some additional mathematical material in Appendix \ref{subs:proofs1}.
\begin{definition}[Type of a sequence \cite{moser:2021}]
	The joint type $P_{\mathbf x, \mathbf y}$ of a pair of length-$n$ sequences $\left(\mathbf x, \mathbf y\right)$ is defined as:
	\begin{IEEEeqnarray*}{c} P_{\mathbf x,\mathbf y}\left(a,b\right)\triangleq \frac{N_o\left(a,b|\mathbf x,\mathbf y\right)}{n}, \quad \forall \quad  \left(a,b\right)\in\mathbb X\times\mathbb Y, \label{types}
	\end{IEEEeqnarray*}
	where by $N_o\left(a,b|\mathbf x,\mathbf y\right)$ we denote the number of times the symbol $\left(a,b\right)$ occurs in $\left(\mathbf x,\mathbf y\right)$.
\end{definition}
We denote as $\mathbf P_n\left(\mathbb X,\mathbb Y\right)$ the set of all joint types with denominator $n$ and with respect to the alphabet $\mathbb X \times \mathbb Y$. We define also the set $\mathbf P^e\left(\mathbb X,\mathbb Y\right)=\cup_{n=1}^{\infty}{\mathbf P_n\left(\mathbb X,\mathbb Y\right)}$. We denote as $\mathbf P\left(\mathbb X,\mathbb Y\right)$ the set of all joint probability mass functions (PMFs) on $\mathbb X \times\mathbb Y$.

\begin{definition}[Matching PMF of a type]  Let a joint PMF $p_{X,Y}\in \mathbf P\left(\mathbb X,\mathbb Y\right) $ and a joint type $P\in \mathbf P^e\left(\mathbb X,\mathbb Y\right) $. We will say that $p_{X,Y}$ is the matching joint PMF of $P$ and we will write $ p_{X,Y}\doteq P$ if 
	$p_{X,Y}\left(a,b\right)=P\left(a,b\right),\quad \forall \left(a,b\right)\in \mathbb X \times \mathbb Y$. 
\end{definition}

\begin{definition}[Conditional type of a sequence \cite{moser:2021}] The conditional type of a sequence $\mathbf y\in \mathbb Y^n$ given a sequence $\mathbf x\in \mathbb X^n$ is defined as:
	\begin{IEEEeqnarray*}{rCl} P_{\mathbf y|\mathbf x}\left(b|a\right)&\triangleq& \frac{N_o\left(a,b|\mathbf x,\mathbf y\right)}{N_o\left(a|\mathbf x\right)}, \quad \forall \quad  \left(a,b\right)\in\mathbb X\times \mathbb Y\quad \\&&\quad \quad \quad \quad \quad \quad \quad \text{with}\quad N_o\left(a|\mathbf x\right)>0. \label{ctypes}
	\end{IEEEeqnarray*}
\end{definition}

We denote as $\mathbf P_n\left(\mathbb Y|\mathbb X\right)$ the set of all conditional types of some length-$n$ sequence $\mathbf y\in\mathbb Y^n$ given some length-$n$ sequence $\mathbf x\in\mathbb X^n$. We define also the set $\mathbf P^e\left(\mathbb Y|\mathbb X\right)=\cup_{n=1}^{\infty}{\mathbf P_n\left(\mathbb Y|\mathbb X\right)}$. We denote as $\mathbf P\left(\mathbb Y|\mathbb X\right)$ the set of all conditional PMFs on $b\in \mathbb Y$ given $a\in \mathbb X$.
\begin{definition}[Matching conditional PMF of a conditional type]  Let a conditional PMF $p_{Y|X}\in \mathbf P\left(\mathbb Y|\mathbb X\right) $ and a conditional type $J\in \mathbf P^e\left(\mathbb Y|\mathbb X\right) $. We will say that $p_{Y|X}$ is the matching conditional PMF of $J$ and we will write $ p_{Y|X}\doteq J$ if 
	$p_{Y|X}\left(b|a\right)=J\left(b|a\right),\quad \forall \left(a,b\right)\in \mathbb X \times \mathbb Y$.
\end{definition}
\begin{definition}[Variational Distance \cite{moser:2021}] The variational distance between two types $Q,N\in \mathbf P^e\left(\mathbb X\right)$ is given by \begin{equation*}V\left(Q,N\right)\triangleq\sum_{a\in \mathbb X}{|Q\left(a\right)-N\left(a\right)|}. \label{variational}\end{equation*}
\end{definition}
\begin{definition}[$\Delta$-neighborhood \cite{mylonakis:2019}]
	The $\Delta$-neighborhood of a type $Q\in \mathbf P^e\left(\mathbb X\right)$ is defined as
	\begin{IEEEeqnarray*}{c} {N_{\Delta}\big(Q\big)}\triangleq \big\{N\in \mathbf P^e\left(\mathbb X\right):V\left(Q,N\right)\leq\Delta\big\}. \label{Deltaneighborhood}
	\end{IEEEeqnarray*}
\end{definition}
\begin{definition}[Pre-image and $\Delta$-pre-image  of $Q$ under $p_{Y|X}$ \cite{serrano:2014}] Let a type $Q\in \mathbf P^e\left(\mathbb 		Y\right)$, a conditional PMF $p_{Y|X}\in \mathbf P\left(\mathbb Y|\mathbb X\right)$ and a conditional type $J\in \mathbf P^e\left(\mathbb Y|\mathbb X\right)$ such that $p_{Y|X}\doteq J$. We define the pre-image and the $\Delta$-pre-image respectively of $Q$ under $p_{Y|X}$ as:
	\begin{IEEEeqnarray*}{rCl}
		\mathbf P_{pim}\left(Q,p_{Y|X}\right)&\triangleq& \left\{N\in P^e\left(\mathbb X\right): \sum_{a\in \mathbb X}N\left(a\right)J\left(\cdot|a\right)=Q\right\}, \label{pim}\\
		\mathbf P^{\Delta}_{pim}\left(Q,p_{Y|X}\right)&\triangleq& \Bigg\{N\in P^e\left(\mathbb X\right):\\&&  \quad \quad \quad  \sum_{a\in \mathbb X}N\left(a\right)J\left(\cdot|a\right)\in N_{\Delta}\left( Q\right)\Bigg\}.
	\end{IEEEeqnarray*}
\end{definition}
\begin{remark}
	It is easily observed that
	\begin{IEEEeqnarray*}{c}\mathbf P^{\Delta}_{pim}\left(Q,p_{Y|X}\right)=\bigcup_{N \in N_{\Delta}\left(Q\right)} \mathbf P_{pim}\left(N,p_{Y|X}\right).  \end{IEEEeqnarray*}
	\label{lem:pim&Deltapim}
\end{remark}
A $\left(2^{nR},n\right)$ communication-coordination code is the protocol which is used for communicating a random message $M$ and at the same time for coordinating with the external observer.

\begin{definition}[Communication-coordination code] The $\left(2^{nR},n\right)$-communication-coordination code for our set-up consists of a message set $\mathbb M=\left\{1,\dots, 2^{nR} \right\}$ and two functions-an encoding function $x^n:\mathbb M
	\rightarrow \mathbb X^n$ and a decoding function
	$\hat{m}:\mathbb Y^n
	\rightarrow \mathbb M.$
	
\end{definition}
\begin{definition}[Induced distribution]
	The induced distribution $\tilde{p}_{Z^n}^s\left(z^n\right)$ is the resulting distribution of the action $Z^n$ of the external observer when $S=s$ and a $\left(2^{nR},n\right)$-communication-coordination code is used.
	\label{def:induced}
\end{definition}

\section{Multiple Interference Coordination}
\label{s:perfect}
In this section, we state and solve the preliminary problem of {\it multiple} interference coordination over a compound channel. In this problem, we ask for reliable communication between the encoder and the decoder and at the same time for the establishment of interference coordination with a different interference type $Q_s$, depending on the exact state of the channel. For the state variable $S$, we assume without loss of generality that it takes values in the set $\mathbb S=\left\{1,\dots,|\mathbb S|\right\}$. By $P_{\mathbf z}$ we denote the type of the sequence $\mathbf z$ which is observed by the external observer.
\begin{definition}[Achievability] We say that a communication rate $R$ is multiply achievable with interference types $\left(Q_1,\dots,Q_{|\mathbb S|}\right)\in \underbrace{ \mathbf P^e\left(\mathbb Z\right)\times \cdots \times \mathbf P^e\left(\mathbb Z\right)}_{|\mathbb S|\quad \text{times}}$ in the compound DMC $p_{Y,Z|X,S}\in \mathbf P\left(\mathbb Y,\mathbb Z|\mathbb X,\mathbb S\right)$ if there exists a sequence of $\left(2^{nR},n\right)$-codes such that 
	
	\begin{IEEEeqnarray}{rCl}&& \lim_{n\to \infty}\max_{s\in\mathbb S}\Pr\left(\hat{M}\neq M|s \quad \text{is the selected channel}\right)= 0,\label{compoundperfecta} \IEEEeqnarraynumspace\\&&V\left(Q_s,P_{\mathbf z}\right)\to 0 \quad \text{in probability}\nonumber\\&& \quad \quad \quad \quad  \quad \quad  \text{if} \quad S=s, \quad  \forall s=1,\dots,|\mathbb S|.\label{compoundperfectb}
	\end{IEEEeqnarray}
\label{def:compoundperfect1}
\end{definition}
\begin{definition}[Communication-multiple interference capacity region]
	The communication-multiple interference region $C$ of the compound DMC  $p_{Y,Z|X,S}\in \mathbf P\left(\mathbb Y,\mathbb Z|\mathbb X,\mathbb S\right)$ is the closure of the set of all rate-interference type tuples $\big(R,Q_1,\dots, Q_{|\mathbb S|}\big)$ that are multiply achievable:
	\begin{multline*}
	C\left(p_{Y,Z|X,S}\right)\triangleq\\
	\mathbf {Cl}\left\{ \,
	\begin{IEEEeqnarraybox}[
	\IEEEeqnarraystrutmode
	\IEEEeqnarraystrutsizeadd{1pt}
	{1pt}][c]{l}
	\big(R,Q_1,\dots, Q_{|\mathbb S|}\big):\\
	R \quad \text{is multiply achievable with interference}\\\text{types}\left(Q_1,\dots, Q_{|\mathbb S|}\right)
	\end{IEEEeqnarraybox}\right\}.
	\end{multline*}
\end{definition}
\begin{theorem}
	The communication-multiple interference capacity region $C$ of the compound DMC $p_{Y,Z|X,S}\in \mathbf P^e\left(\mathbb Y,\mathbb Z|\mathbb X,\mathbb S\right)$ is the set
	
	\begin{multline*}
	C\left(p_{Y,Z|X,S}\right)=\\
	\mathbf {Cl}\left\{ \,
	\begin{IEEEeqnarraybox}[
	\IEEEeqnarraystrutmode
	\IEEEeqnarraystrutsizeadd{1pt}
	{1pt}][c]{l}
	\big(R,Q_1,\dots,Q_{|\mathbb S|}\big):\\
	R\leq \max_{N\in \bigcap_{s=1}^{|\mathbb S|} \mathbf P_{\text{pim}}\left(Q_s,p_{Z|X,S}\left(\cdot|\cdot,S=s\right)\right)}\min_{s\in\mathbb S} I\left(NJ_s\right)
	\end{IEEEeqnarraybox}\right\},
	\end{multline*}
	if $\bigcap_{s=1}^{|\mathbb S|} \mathbf P_{\text{pim}}\left(Q_s,p_{Z|X,S}\left(\cdot|\cdot,S=s\right)\right)\neq \emptyset$ and where $p_{Y|X,S}\in \mathbf P\left(\mathbb Y|\mathbb X,\mathbb S\right)$ is the marginal conditional PMF of $p_{Y,Z|X,S}$ and $J_s\in \mathbf P^e\left(\mathbb Y|\mathbb X\right)$ is a conditional type such that $p_{Y|X,S}\left(\cdot|\cdot,s\right)\doteq J_s$. 
\label{th:compoundperfect}	
\end{theorem}
\begin{IEEEproof}See Section \ref{subs:proofs2}.
\end{IEEEproof}
\section{Adaptive Interference Coordination}
\label{s:imperfect}
In this section, we state and solve the problem of {\it adaptive} interference coordination over a compound channel. In this problem, we ask for reliable communication between the encoder and the decoder and at the same time for the establishment of interference coordination of different precision $\Delta_s$ with the same interference type $Q$, depending on the exact state $s$ of the channel. By $\mathbb E_{\tilde p^s_{Z^n}}\left\{\cdot \right\}$ we denote the expected value of a quantity with respect to the induced distribution of the coordination code (see Definition \ref{def:induced}).
\begin{definition}[Achievability] We say that a communication rate $R$ is $\Delta_1\cdots\Delta_{|\mathbb S|}$-achievable with interference type $Q\in \mathbf P^e\left(\mathbb Z\right)$ in the compound DMC  $p_{Y,Z|X,S}\in \mathbf P\left(\mathbb Y,\mathbb Z|\mathbb X,\mathbb S\right)$ if for all $\epsilon>0$, there exists an $N$ such that for all $n>N$, there exist $\left(2^{nR},n\right)$-codes such that 
	
	\begin{IEEEeqnarray*}{rCl}&&\max_{s\in\mathbb S}\Pr\left(\hat{M}\neq M|s \quad \text{is the selected channel}\right)<\epsilon,\\&&\mathbb E_{\tilde p^s_{Z^n}}\big\{V\left(Q,P_{\mathbf z}\right)\big\}\leq \Delta_s \quad \text{if} \quad S=s, \quad \forall s=1,\dots,|\mathbb S|.
    \end{IEEEeqnarray*}
	\end{definition}
\begin{definition}[Communication-adaptive interference capacity region]
	The communication-adaptive interference region $R^I$ of the compound DMC $p_{Y,Z|X,S}\in \mathbf P\left(\mathbb Y,\mathbb Z|\mathbb X,\mathbb S\right)$ with the interference type $Q\in \mathbf P^e\left(\mathbb Z\right)$  is the closure of the set of all rate-distortion tuples $\big(R,\Delta_1,\dots,\Delta_{|\mathbb S|}\big)$ such that $R$ to be $\Delta_1\cdots\Delta_{|\mathbb S|}$-achievable with interference type $Q$:
	\begin{equation*}
	R^I\left(Q\right)\triangleq
	\mathbf {Cl}\left\{ \,
	\begin{IEEEeqnarraybox}[
	\IEEEeqnarraystrutmode
	\IEEEeqnarraystrutsizeadd{1pt}
	{1pt}][c]{l}
	\big(R,\Delta_1,\dots,\Delta_{|\mathbb S|}\big):
	R \quad \text{is $\Delta_1\cdots\Delta_{|\mathbb S|}$}\\\text{-achievable with interference type $Q$}
	\end{IEEEeqnarraybox}\right\}.
	\end{equation*}

\end{definition}
\begin{theorem}
	For every compound DMC $p_{Y,Z|X,S} \in \mathbf P\left(\mathbb Y,\mathbb Z|\mathbb X,\mathbb S\right)$ and every interference type $Q\in \mathbf P^e\left(\mathbb Z\right)$:

\begin{multline*}
R^I\big(Q\big)\quad=
\left.
\left\{ \,
\begin{IEEEeqnarraybox}[
\IEEEeqnarraystrutmode
\IEEEeqnarraystrutsizeadd{2pt}
{2pt}][c]{l}
\left(R,\Delta_1,\dots,\Delta_{|\mathbb S|}\right):\\
\left(R,Q_1,\dots,Q_{|\mathbb S|}\right)\in C\left(p_{Y,Z|X,S}\right)\\\text{for some $\left(Q_1,\dots,Q_{|\mathbb S|}\right)$ such that}\\\text{ $Q_s \in  N_{\Delta_s}\left(Q\right),\quad \forall s=1,\dots,|\mathbb S|$}
\end{IEEEeqnarraybox}\right\}.
\right.
\end{multline*}
\label{th:compoundperfectimperfect}		
\end{theorem}
\begin{IEEEproof}See Section \ref{subs:proofs3}.
\end{IEEEproof}

\begin{theorem}
	The communication-adaptive interference capacity region $R^I$ of the compound DMC $p_{Y,Z|X,S}\in \mathbf P^e\left(\mathbb Y,\mathbb Z|\mathbb X,\mathbb S\right)$ with the interference type $Q\in \mathbf P^e\left(\mathbb Z\right)$ is the set
	
	\begin{multline*}
	R^I\left(Q\right)=\\
	\mathbf {Cl}\left\{ \,
	\begin{IEEEeqnarraybox}[
	\IEEEeqnarraystrutmode
	\IEEEeqnarraystrutsizeadd{1pt}
	{1pt}][c]{l}
	\big(R,\Delta_1,\dots,\Delta_{|\mathbb S|}\big):\\
	R\leq \max_{N\in \bigcap_{s=1}^{|\mathbb S|} \mathbf P^{\Delta_s}_{\text{pim}}\left(Q,p_{Z|X,S}\left(\cdot|\cdot,S=s\right)\right)}\min_{s\in\mathbb S} I\left(NJ_s\right)
	\end{IEEEeqnarraybox}\right\}.
	\end{multline*}
	if $ \bigcap_{s=1}^{|\mathbb S|} \mathbf P^{\Delta_s}_{\text{pim}}\left(Q,p_{Z|X,S}\left(\cdot|\cdot,S=s\right)\right)\neq \emptyset$ and where $p_{Y|X,S}\in \mathbf P\left(\mathbb Y|\mathbb X,\mathbb S\right)$ is the marginal conditional PMF of $p_{Y,Z|X,S}$ and $J_s\in \mathbf P^e\left(\mathbb Y|\mathbb X\right)$ is a conditional type such that $p_{Y|X,S}\left(\cdot|\cdot,s\right)\doteq J_s$.
	\label{th:compoundimperfect}
\end{theorem}
\begin{IEEEproof}From Theorem \ref{th:compoundperfectimperfect} and Theorem \ref{th:compoundperfect} we get:
\begin{multline*}
	R^I\left(Q\right)=\\
	\mathbf {Cl}\left\{ \,
	\begin{IEEEeqnarraybox}[
	\IEEEeqnarraystrutmode
	\IEEEeqnarraystrutsizeadd{1pt}
	{1pt}][c]{l}
	\big(R,\Delta_1,\dots,\Delta_{|\mathbb S|}\big):\\
	R\leq \\ \max_{\substack{\left(Q_1,\dots,Q_{|\mathbb S|}\right)\\\in N_{\Delta_1}\left(Q\right)\times\\ \dots \times N_{\Delta_{|\mathbb S|}}\left(Q\right)}} \max_{\substack{N\in\\ \bigcap_{s=1}^{|\mathbb S|}\\ \mathbf P_{\text{pim}}\left(Q_s,p_{Z|X,S}\left(\cdot|\cdot,S=s\right)\right)}}\min_{s\in\mathbb S} I\left(NJ_s\right)
	\end{IEEEeqnarraybox}\right\},
	\end{multline*}
 which also can be written as
 \begin{multline*}
 R^I\left(Q\right)=\\
 \mathbf {Cl}\left\{ \,
 \begin{IEEEeqnarraybox}[
 \IEEEeqnarraystrutmode
 \IEEEeqnarraystrutsizeadd{1pt}
 {1pt}][c]{l}
 \big(R,\Delta_1,\dots,\Delta_{|\mathbb S|}\big):
 R\leq \\ \max_{\substack{N\in \bigcup_{\left(Q_1,\dots,Q_{|\mathbb S|}\right)\in N_{\Delta_1}\left(Q\right)\times \dots N_{\Delta_{|\mathbb S|}}\left(Q\right)}\\ \bigcap_{s=1}^{|\mathbb S|} \mathbf P_{\text{pim}}\left(Q_s,p_{Z|X,S}\left(\cdot|\cdot,S=s\right)\right)}}\min_{s\in\mathbb S} I\left(NJ_s\right)
 \end{IEEEeqnarraybox}\right\},
 \end{multline*}
 and by changing the order of the union and the intersection
 \begin{equation*}
R^I\left(Q\right)=
\mathbf {Cl}\left\{ \,
\begin{IEEEeqnarraybox}[
\IEEEeqnarraystrutmode
\IEEEeqnarraystrutsizeadd{1pt}
{1pt}][c]{l}
\big(R,\Delta_1,\dots,\Delta_{|\mathbb S|}\big):
R\leq \\ \max_{\substack{N\in \bigcap_{s=1}^{|\mathbb S|}\bigcup_{Q_s\in N_{\Delta_{s}}\left(Q\right)}\\  \mathbf P_{\text{pim}}\left(Q_s,p_{Z|X,S}\left(\cdot|\cdot,S=s\right)\right)}}\min_{s\in\mathbb S} I\left(NJ_s\right)
\end{IEEEeqnarraybox}\right\},
\end{equation*}
and finally by Remark \ref{lem:pim&Deltapim} we get the desired. This completes the proof.

\end{IEEEproof}
\appendices

\section{Proof of Theorem \ref{th:compoundperfect}}
\label{subs:proofs2}
\paragraph*{\bf Achievability}
\begin{itemize}
	\item {Setup:} We assume that $\epsilon_1>0$ is given. Then we fix some rate $R$, some blocklength $n$, and some $\epsilon>0$.
	\item{Message set:} We assume that the source randomly picks a message $\mathbb M=\left\{1,\dots,\lceil 2^{n\left(R-\epsilon_1\right)} \rceil\right\}$ according to a uniform distribution $\Pr\left(M=m\right)=\frac{1}{\lceil 2^{n\left(R-\epsilon_1\right)} \rceil}$.
	\item {Codebook:} Fix a codebook $C\in \mathbb C $ of $\lceil e^{n\left(R+\epsilon_1\right)} \rceil$ length-$n$ codewords $\mathbf x^C\left(m\right), m=1,\dots, \lceil e^{n\left(R-\epsilon_1\right)} \rceil$.
	\item {Encoder Design:} The encoder maps each message $m\in \mathbb M$ to the sequence $\mathbf x^C\left(m\right)$.
	\item {Decoder Design:} The decoder receives a sequence $\mathbf y$ and tries to guess which message has been sent. It searches for an $\mathbf x^C\left(\hat{m}\right)$ such that $\left(\mathbf x^C\left(\hat{m}\right),\mathbf y\right)\in \mathbb{A}_{\epsilon}^{\ast\left(n\right)}\left(NJ_s\right)$ for some $s\in \mathbb S$. If it finds exactly one $\hat{m}$, it outputs $\hat{m}$. Otherwise, it declares an error.
	\item {Performance Analysis:} We do not try to analyze one particular codebook, but instead the average over all possible codebooks. In particular, we define the PMF $p_X\doteq N$ where $ N\in \bigcap_{s=1}^{|\mathbb S|} \mathbf P_{\text{pim}}\left(Q_s,p_{Z|X,S}\left(\cdot|\cdot,S=s\right)\right) $ and we generate the $\lceil 2^{n\left(R-\epsilon_1\right)} \rceil$ length-$n$ codewords $\mathbf x^C\left(m\right)$ of the code by choosing each of the $n\lceil 2^{n\left(R-\epsilon_1\right)} \rceil$ symbols $\mathbf x_k^C\left(m\right)$  independently at random according to $p_X$. If the average probability of error goes to zero as $n$ tends to infinity, then we conclude that is should exist at least one sequence of coordination codes with probability of error tending to zero. We break our analysis in two parts: (a) the decoding part (b) the interference coordination part. We denote as $\mathbb E_{\mathbb C}\left\{\cdot\right\}$ the expected value of a quantity with respect to the distribution over the codes and as $\mathbbm{1}_C\left(\cdot\right)$ a function which becomes one if the claim which includes is true for this specific code $C$. \par We begin with the decoding part. Due to symmetry, we can assume without loss of generality that $M=1$. For the error analysis, we distinguish two different types of errors: (a) $\left\{\left(\mathbf x^C\left(1\right),\mathbf y\right)\notin \mathbb{A}_{\epsilon}^{\ast\left(n\right)}\left(NJ_s\right)\quad \forall s\right\}$, (b) $\Big\{\exists m\neq 1, \exists s:\left(\mathbf x^C\left(m\right),\mathbf y\right)\in \mathbb{A}_{\epsilon}^{\ast\left(n\right)}\left(NJ_s\right)\Big\}$.
	First, the error of type (a). We define $\epsilon^\prime\triangleq \frac{\epsilon}{2|\mathbb Y|}$ :
	\begin{IEEEeqnarray}{rCl}&&\mathbb E_{\mathbb C}\big\{\max_{s\in\mathbb S}\Pr\left(\left(\mathbf x^C\left(1\right),\mathbf y\right)\notin \mathbb{A}_{\epsilon}^{\ast\left(n\right)}\left(NJ_s\right)\big|\right.\nonumber\\&&\left.\quad \quad \quad \quad \quad \quad \quad \quad  s \quad \text{is the selected channel}\right)\big\}\nonumber\\
		&=&\sum_{C\in \mathbb C}\left(\prod_{m}\Pr\left(p_X,\mathbf x^C\left(m\right)\right)\right)\cdot\nonumber\\&&\quad \quad \quad \cdot\max_{s\in\mathbb S}\Bigg(\sum_{\mathbf y\in \mathbb X^n}\Pr\bigg(p_{Y|X,S}\big(\cdot|\cdot,S=s\big),\mathbf y\Big|\mathbf x^C\left(1\right)\bigg)\cdot\nonumber\end{IEEEeqnarray}\begin{IEEEeqnarray}{rCl}&&\quad \quad \quad \quad \quad \quad  \quad \cdot \mathbbm{1}_C\Big(\big(\mathbf x^C\left(1\right), \mathbf y\big)\notin  \mathbb{A}_{\epsilon}^{\ast\left(n\right)}\left(NJ_s\right)\Big)\Bigg)\nonumber\\&=&\sum_{C\in \mathbb C}\left(\prod_{m}\Pr\left(p_X,\mathbf x^C\left(m\right)\right)\right)\cdot\nonumber\\&&\quad \quad \quad\cdot\max_{s\in \mathbb S}\Bigg(\sum_{\mathbf y\in \mathbb X^n}\Pr\bigg(p_{Y|X,S}\big(\cdot|\cdot,S=s\big),\mathbf y\Big|\mathbf x^C\left(1\right)\bigg)\nonumber\\&&\quad \quad \quad \quad \quad \quad \quad \bigg(1-\mathbbm{1}_C\Big(\big(\mathbf x^C\left(1\right), \mathbf y\big)\in  \mathbb{A}_{\epsilon}^{\ast\left(n\right)}\left(NJ_s\right)\Big)\bigg)\Bigg)\nonumber\\&=& \sum_{C\in \mathbb C}\left(\prod_{m}\Pr\left(p_X,\mathbf x^C\left(m\right)\right)\right)\cdot\nonumber\\&&\quad \quad \quad \cdot\max_{s\in\mathbb S}\Bigg(\sum_{\mathbf y\in \mathbb X^n}\Pr\bigg(p_{Y|X,S}\big(\cdot|\cdot,S=s\big),\mathbf y\Big|\mathbf x^C\left(1\right)\bigg)\Bigg)\nonumber\\&&-\sum_{C\in \mathbb C}\left(\prod_{m}\Pr\left(p_X,\mathbf x^C\left(m\right)\right)\right)\cdot\nonumber\\&&\quad \quad \quad  \cdot \max_{s\in \mathbb S}\Bigg(\sum_{\mathbf y\in \mathbb X^n}\Pr\bigg(p_{Y|X,S}\big(\cdot|\cdot,S=s\big),\mathbf y\Big|\mathbf x^C\left(1\right)\bigg)\cdot\nonumber\\&&\quad \quad \quad \quad \quad \quad \quad \quad \cdot\mathbbm{1}_C\Big(\big(\mathbf x^C\left(1\right), \mathbf y\big)\in  \mathbb{A}_{\epsilon}^{\ast\left(n\right)}\left(NJ_s\right)\Big)\Bigg)\nonumber\\&=& \sum_{C\in \mathbb C}\left(\prod_{m}\Pr\left(p_X,\mathbf x^C\left(m\right)\right)\right)\cdot\nonumber\\&&\quad \quad \quad \cdot\max_{s\in\mathbb S}\Bigg(\sum_{\mathbf y\in \mathbb X^n}\Pr\bigg(p_{Y|X,S}\big(\cdot|\cdot,S=s\big),\mathbf y\Big|\mathbf x^C\left(1\right)\bigg)\nonumber\\&&-\sum_{C\in \mathbb C}\left(\prod_{m}\Pr\left(p_X,\mathbf x^C\left(m\right)\right)\right)\cdot\nonumber\\&&\quad \quad \quad \quad \cdot\max_{s\in \mathbb S}\Bigg(\sum_{\mathbf y\in \mathbb X^n}\Pr\bigg(p_{Y|X,S}\big(\cdot|\cdot,S=s\big),\mathbf y\Big|\mathbf x^C\left(1\right)\bigg)\cdot\nonumber\\&& \quad \quad \quad \quad \quad \quad \quad \quad \cdot\mathbbm{1}_C\Big(\mathbf x^C\left(1\right)\in \mathbb{A}_{\epsilon}^{\ast\left(n\right)}\left(N\right)\Big)\cdot\nonumber\\&&\quad \quad \quad \quad \quad \quad \quad \quad \cdot\mathbbm{1}_C\Big(\mathbf y \in \mathbb{A}_{\epsilon}^{\ast\left(n\right)}\left(NJ_s|\mathbf x_C\left(1\right)\right)\Big)\Bigg)\label{achievabilitya1}\\  &\leq& \sum_{C\in \mathbb C}\left(\prod_{m}\Pr\left(p_X,\mathbf x^C\left(m\right)\right)\right)\cdot\nonumber\\&&\quad \quad \quad \cdot\max_{s\in\mathbb S}\Bigg(\sum_{\mathbf y\in \mathbb X^n}\Pr\bigg(p_{Y|X,S}\big(\cdot|\cdot,S=s\big),\mathbf y\Big|\mathbf x^C\left(1\right)\bigg)\Bigg)\nonumber\\&&-\sum_{C\in \mathbb C}\left(\prod_{m}\Pr\left(p_X,\mathbf x^C\left(m\right)\right)\right)\cdot\nonumber\\&&\quad \quad \quad \quad \cdot \max_{s\in \mathbb S}\Bigg(\sum_{\mathbf y\in \mathbb X^n}\Pr\bigg(p_{Y|X,S}\big(\cdot|\cdot,S=s\big),\mathbf y\Big|\mathbf x^C\left(1\right)\bigg)\cdot\nonumber \\&& \quad \quad \quad \quad \quad \quad \quad \quad \cdot\mathbbm{1}_C\Big(\mathbf x^C\left(1\right)\in \mathbb{A}_{\epsilon^\prime}^{\ast\left(n\right)}\left(N\right)\Big)\cdot\nonumber\\&&\quad \quad \quad \quad \quad \quad \quad \quad \cdot\mathbbm{1}_C\Big(\mathbf y \in \mathbb{A}_{\epsilon}^{\ast\left(n\right)}\left(NJ_s|\mathbf x_C\left(1\right)\right)\Big)\Bigg)\label{achievabilitya2}\end{IEEEeqnarray}\begin{IEEEeqnarray}{rCl} &=&1-\nonumber\\&&\sum_{C\in \mathbb C}\left(\prod_{m}\Pr\left(p_X,\mathbf x^C\left(m\right)\right)\right)\mathbbm{1}_C\Big(\mathbf x^C\left(1\right)\in \mathbb{A}_{\epsilon^\prime}^{\ast\left(n\right)}\left(N\right)\Big)\cdot\nonumber\\&&\quad \quad \quad \cdot\max_{s\in \mathbb S}\Bigg(\sum_{\mathbf y\in \mathbb X^n}\Pr\bigg(p_{Y|X,S}\big(\cdot|\cdot,S=s\big),\mathbf y\Big|\mathbf x^C\left(1\right)\bigg)\cdot\nonumber\\&&\quad \quad \quad \quad \quad \quad \quad \cdot \mathbbm{1}_C\Big(\mathbf y \in \mathbb{A}_{\epsilon}^{\ast\left(n\right)}\left(NJ_s|\mathbf x_C\left(1\right)\right)\Big)\Bigg)\nonumber\\&=&1-\nonumber\\&&\sum_{C\in \mathbb C}\left(\prod_{m}\Pr\left(p_X,\mathbf x^C\left(m\right)\right)\right)\mathbbm{1}_C\Big(\mathbf x^C\left(1\right)\in \mathbb{A}_{\epsilon^\prime}^{\ast\left(n\right)}\left(N\right)\Big)\cdot\nonumber\\&& \cdot\max_{s\in\mathbb S}\Bigg(\underbrace{\Pr\bigg(p_{Y|X,S}\big(\cdot|\cdot,S=s\big),\mathbb{A}_{\epsilon}^{\ast\left(n\right)}\left(NJ_s\right) \Big|\mathbf x^C\left(1\right)\bigg)}_{>1-\delta_t\left(n,\frac{\epsilon}{2},\mathbb X\times \mathbb Y\right)}\Bigg)\nonumber\\&\leq&1-\bigg(1-\delta_t\left(n,\frac{\epsilon}{2},\mathbb X\times \mathbb Y\right)\bigg)\cdot\nonumber\\&&\cdot\underbrace{\sum_{C\in \mathbb C}\left(\prod_{m}\Pr\left(p_X,\mathbf x^C\left(m\right)\right)\right)\mathbbm{1}_C\Big(\mathbf x^C\left(1\right)\in \mathbb{A}_{\epsilon^\prime}^{\ast\left(n\right)}\left(N\right)\Big)}_{=\Pr\left(p_X,\mathbb{A}_{\epsilon^\prime}^{\ast\left(n\right)}\left(N\right)\right)\leq 1}\IEEEeqnarraynumspace\label{achievabilitya3}\\&\leq&\delta_t\left(n,\frac{\epsilon}{2},\mathbb X\times \mathbb Y\right)\nonumber.
	\end{IEEEeqnarray}
	Here, in \eqref{achievabilitya1} we apply Lemma \ref{lem:cJstsets} in  Section \ref{subs:typicalsets}, in \eqref{achievabilitya2} we substitute $\mathbbm{1}_C\Big(\mathbf x^C\left(1\right)\in \mathbb{A}_{\epsilon}^{\ast\left(n\right)}\left(N\right)\Big)$ by the smaller $\mathbbm{1}_C\Big(\mathbf x^C\left(1\right)\in \mathbb{A}_{\epsilon^\prime}^{\ast\left(n\right)}\left(N\right)\Big)$ and in \eqref{achievabilitya3} we apply Theorem \ref{th:cstsets} in Section \ref{subs:typicalsets}.
	We continue with the error of type (b):
	\begin{IEEEeqnarray}{rCl}&&\mathbb E_{\mathbb C}\big\{\max_{s\in\mathbb S}\Pr\left(\left(\mathbf x^C\left(m\right),\mathbf y\right)\in \mathbb{A}_{\epsilon}^{\ast\left(n\right)}\left(NJ_s\right)\quad \text{for some}\right.\nonumber\\&&\left.   \quad  m \neq 1 \quad \text{and for some}\quad s^{\prime}|s \quad \text{is the selected channel}\right)\big\}\nonumber\\
		&=&\sum_{C\in \mathbb C}\left(\prod_{m}\Pr\left(p_X,\mathbf x^C\left(m\right)\right)\right)\cdot\nonumber\\&& \quad\cdot\max_{s\in\mathbb S}\Bigg(\sum_{\mathbf y\in \mathbb X^n}\Pr\bigg(p_{Y|X,S}\big(\cdot|\cdot,S=s\big),\mathbf y\Big|\mathbf x^C\left(1\right)\bigg)\cdot\nonumber\\&&\quad  \quad \quad  \cdot\mathbbm{1}_C\Big(\exists m\neq1,\exists s^{\prime}:\big(\mathbf x^C\left(m\right), \mathbf y\big)\in  \mathbb{A}_{\epsilon}^{\ast\left(n\right)}\left(NJ_{s^{\prime}}\right)\Big)\Bigg)\nonumber\\&=&\sum_{C\in \mathbb C}\left(\prod_{m\neq 1}\Pr\left(p_X,\mathbf x^C\left(m\right)\right)\right)\Pr\left(p_X,\mathbf x^C\left(1\right)\right)\cdot\nonumber\\&&\quad  \quad \cdot\max_{s\in\mathbb S}\Bigg(\sum_{\mathbf y\in \mathbb X^n}\Pr\bigg(p_{Y|X,S}\big(\cdot|\cdot,S=s\big),\mathbf y\Big|\mathbf x^C\left(1\right)\bigg)\cdot\nonumber\\&&   \quad \quad \quad  \cdot \mathbbm{1}_C\Big(\exists m\neq1,\exists s^{\prime}:\big(\mathbf x^C\left(m\right), \mathbf y\big)\in  \mathbb{A}_{\epsilon}^{\ast\left(n\right)}\left(NJ_{s^{\prime}}\right)\Big)\Bigg)\nonumber\end{IEEEeqnarray}\begin{IEEEeqnarray}{rCl}&=&\sum_{C\in \mathbb C}\left(\prod_{m\neq 1}\Pr\left(p_X,\mathbf x^C\left(m\right)\right)\right)\cdot\nonumber\\&&  \quad \cdot\max_{s\in\mathbb S}\Bigg(\sum_{\mathbf y\in \mathbb X^n}\Pr\left(p_X,\mathbf x^C\left(1\right)\right)\cdot\nonumber\\&& \quad \quad \quad \cdot \Pr\bigg(p_{Y|X,S}\big(\cdot|\cdot,S=s\big),\mathbf y\Big|\mathbf x^C\left(1\right)\bigg)\cdot\nonumber\\&&   \quad \quad \quad  \cdot\mathbbm{1}_C\Big(\exists m\neq1,\exists s^{\prime}:\big(\mathbf x^C\left(m\right), \mathbf y\big)\in  \mathbb{A}_{\epsilon}^{\ast\left(n\right)}\left(NJ_{s^{\prime}}\right)\Big)\Bigg)\nonumber\\&=&\sum_{C\in \mathbb C}\left(\prod_{m\neq 1}\Pr\left(p_X,\mathbf x^C\left(m\right)\right)\right)\cdot\nonumber\\&& \quad \quad \cdot\max_{s\in\mathbb S}\Bigg(\sum_{\mathbf y\in \mathbb X^n} \Pr\bigg(p_{X,Y|S}\big(\cdot,\cdot|S=s\big),\Big(\mathbf x^C\left(1\right),\mathbf y\Big) \bigg)\cdot\nonumber\\&&  \quad \quad \cdot  \mathbbm{1}_C\Big(\exists m\neq1,\exists s^{\prime}:\big(\mathbf x^C\left(m\right), \mathbf y\big)\in  \mathbb{A}_{\epsilon}^{\ast\left(n\right)}\left(NJ_{s^{\prime}}\right)\Big)\Bigg)\nonumber\\&\leq&\sum_{C\in \mathbb C}\left(\prod_{m\neq 1}\Pr\left(p_X,\mathbf x^C\left(m\right)\right)\right)\cdot\nonumber\\&&   \cdot\max_{s\in\mathbb S}\Bigg(\sum_{\mathbf y\in \mathbb X^n} \Pr\bigg(p_{Y|S}\big(\cdot|S=s\big),\mathbf y \bigg)\cdot\nonumber\\&&    \cdot \mathbbm{1}_C\Big(\exists m\neq1,\exists s^{\prime}:\big(\mathbf x^C\left(m\right), \mathbf y\big)\in  \mathbb{A}_{\epsilon}^{\ast\left(n\right)}\left(NJ_{s^\prime}\right)\Big)\Bigg)\label{achievabilityb1}\\&=&\sum_{C\in \mathbb C}\left(\prod_{m\neq 1}\Pr\left(p_X,\mathbf x^C\left(m\right)\right)\right)\cdot\nonumber\\&&\quad \quad \quad \cdot\max_{s\in\mathbb S}\Bigg(\sum_{\mathbf y\in \mathbb X^n} \Pr\bigg(p_{Y|S}\big(\cdot|S=s\big),\mathbf y \bigg)\cdot\nonumber\\ &&  \quad \quad \quad \quad \quad \cdot\mathbbm{1}_C\Big(\exists m\neq1,\exists s^{\prime}:\mathbf x^C\left(m\right)\in  \mathbb{A}_{\epsilon}^{\ast\left(n\right)}\left(N\right)\nonumber\\&& \quad \quad \quad \quad \quad \quad \quad\wedge \mathbf y\in \mathbb{A}_{\epsilon}^{\ast\left(n\right)}\left(NJ_{s^{\prime}}|\mathbf x^C\left(m\right)\right) \Big)\Bigg)\label{achievabilityb2}\\&\leq& \sum_{C\in \mathbb C}\left(\prod_{m\neq 1}\Pr\left(p_X,\mathbf x^C\left(m\right)\right)\right)\cdot\nonumber\\&&\quad \quad \quad \cdot \max_{s\in\mathbb S}\Bigg(\sum_{m\neq 1}\sum_{s^{\prime}}\sum_{\mathbf y\in \mathbb X^n} \Pr\bigg(p_{Y|S}\big(\cdot|S=s\big),\mathbf y \bigg)\cdot\nonumber\\  && \quad \quad \quad \quad \quad \quad \quad \cdot\mathbbm{1}_C\Big(\mathbf y\in \mathbb{A}_{\epsilon}^{\ast\left(n\right)}\left(NJ_{s^{\prime}}|\mathbf x^C\left(m\right)\right) \Big)\Bigg)\label{achievabilityb3}\\
		&=& \sum_{C\in \mathbb C} \left(\prod_{m\neq 1}\Pr\left(p_X,\mathbf x^C\left(m\right)\right)\right)\max_{s\in\mathbb S}\sum_{m\neq 1}\sum_{s^{\prime}}\nonumber\\ &&\quad \quad \quad \quad \Pr\bigg(p_{Y|S}\big(\cdot|S=s\big),\mathbf \mathbb{A}_{\epsilon}^{\ast\left(n\right)}\left(NJ_{s^{\prime}}|\mathbf x^C\left(m\right)\right)\bigg)
		\label{achievabilityb4}\IEEEeqnarraynumspace\end{IEEEeqnarray}\begin{IEEEeqnarray}{rCl}&<&\sum_{C\in \mathbb C}\left(\prod_{m\neq 1}\Pr\left(p_X,\mathbf x^C\left(m\right)\right)\right)\cdot\nonumber\\&&\cdot \max_{s\in\mathbb S}\Bigg(\sum_{m\neq 1}
		\sum_{s^{\prime} }\text{exp}\bigg(-n\Big(I\left(NJ_{s^{\prime}}\right)\nonumber\\&&\quad \quad \quad \quad \quad \quad \quad \quad \quad \quad \quad +D\left(M_{s^{\prime}}||p_{Y|S}\left(\cdot|S=s\right)\right)\nonumber\\&&\quad \quad \quad \quad \quad \quad \quad \quad \quad \quad \quad -\epsilon_m\big(p_{X}p_{Y|X,S}\left(\cdot|\cdot, S=s^{\prime}\right)\big)\nonumber\\&&\quad \quad \quad \quad \quad \quad  \quad \quad \quad \quad \quad -\epsilon_m\big(p_{Y|S}\left(\cdot|S=s\right)\big)\Big)\bigg)\Bigg)\label{achievabilityb5}\IEEEeqnarraynumspace\\&\leq&\sum_{C\in \mathbb C}\left(\prod_{m\neq 1}\Pr\left(p_X,\mathbf x^C\left(m\right)\right)\right)\cdot\nonumber\\&& \cdot\max_{s\in\mathbb S}\Bigg(\sum_{m\neq 1}
		\sum_{s^{\prime} }\text{exp}\bigg(-n\Big(I\left(NJ_{s^{\prime}}\right)\nonumber\\&&\quad \quad\quad \quad \quad \quad \quad \quad \quad  -\epsilon_m\big(p_{X}p_{Y|X,S}\left(\cdot|\cdot, S=s^{\prime}\right)\big)\nonumber\\&&\quad \quad \quad \quad \quad \quad \quad \quad  \quad  -\epsilon_m\big(p_{Y|S}\left(\cdot|S=s\right)\big)\Big)\bigg)\Bigg)\label{achievabilityb6}\\&\leq&\sum_{C\in \mathbb C}\left(\prod_{m\neq 1}\Pr\left(p_X,\mathbf x^C\left(m\right)\right)\right)\cdot\nonumber\\&&  \cdot\max_{s\in\mathbb S}\Bigg(\lceil e^{n\left(R-\epsilon_1\right)} \rceil \bigg(
		\sum_{s^{\prime}} \text{exp}\Big(-n\big(I\left(NJ_{s^{\prime}}\right)\nonumber\\&&\quad \quad \quad \quad \quad \quad \quad \quad \quad \quad  -\epsilon_m\left(p_{X}p_{Y|X,S}\left(\cdot|\cdot, S=s^{\prime}\right)\right)\nonumber\\&&\quad \quad \quad \quad \quad \quad \quad \quad \quad \quad -\epsilon_m\left(p_{Y|S}\left(\cdot|S=s\right)\right)\big)\Big)\bigg)\Bigg)\nonumber\\&=&\sum_{C\in \mathbb C}\left(\prod_{m\neq 1}\Pr\left(p_X,\mathbf x^C\left(m\right)\right)\right)\cdot \nonumber\\&&\quad \quad \quad \cdot\max_{s\in\mathbb S}\bigg(
		\sum_{s^{\prime}}e^{-n\big(I\left(NJ_{s^{\prime}}\right)-R+\epsilon_1-\delta_{s,s^{\prime}}\big)}\bigg)\nonumber\\&\leq&\sum_{C\in \mathbb C}\left(\prod_{m\neq 1}\Pr\left(p_X,\mathbf x^C\left(m\right)\right)\right)\cdot\nonumber\\&&\quad  \quad  \cdot \Bigg(
		\sum_{s^{\prime}} e^{-n\big(I\left(NJ_{s^{\prime}}\right)+\min_{s\in \mathbb S}\big(-\delta_{s,s^{\prime}}\big)-R+\epsilon_1\big)}\Bigg)\nonumber\\&\leq&\sum_{C\in \mathbb C}\left(\prod_{m\neq 1}\Pr\left(p_X,\mathbf x^C\left(m\right)\right)\right)\cdot\nonumber\\&&\quad \quad \quad \cdot\Bigg(
		\sum_{s^{\prime}}e^{-n\big(I\left(NJ_{s^{\prime}}\right)+\delta^{\prime}-R+\epsilon_1\big)}\Bigg)\nonumber\\&=&\sum_{s^{\prime}}e^{-n\big(I\left(NJ_{s^{\prime}}\right)-R+\epsilon_1+\delta^{\prime}\big)}\cdot\nonumber\\&&\cdot\underbrace{\sum_{C\in \mathbb C}\left(\prod_{m\neq 1}\Pr\left(p_X,\mathbf x^C\left(m\right)\right)\right)}_{\leq 1}\nonumber\\&\leq&\sum_{s^{\prime}}e^{-n\big(I\left(NJ_{s^{\prime}}\right)-R+\epsilon_1+\delta^{\prime}\big)}\nonumber,\end{IEEEeqnarray}
	where $\delta_{s,s^{\prime}}$ account for the rounding mistake and includes the $\epsilon_m$-terms and $\delta^{\prime}\triangleq \min_{s\in\mathbb S}\left(-\delta_{s,s^{\prime}}\right)$. 
	Here, in \eqref{achievabilityb1} we apply the simple fact that $\Pr\bigg(p_{X,Y|S}\big(\cdot,\cdot|S=s\big),\Big(\mathbf x^C\left(1\right),\mathbf y\Big) \bigg)\leq \Pr\bigg(p_{Y|S}\big(\cdot|S=s\big),\mathbf y \bigg)$, in \eqref{achievabilityb2} we apply Lemma \ref{lem:cJstsets} in Section \ref{subs:typicalsets}, in \eqref{achievabilityb3} we enlarge the set inside the function $\mathbbm{1}_C\left(\cdot\right)$, in \eqref{achievabilityb4} we substitute $\sum_{\mathbf y\in \mathbb X^n} \Pr\bigg(p_{Y|S}\big(\cdot|S=s\big),\mathbf y \bigg)\nonumber \mathbbm{1}_C\Big(\mathbf y\in \mathbb{A}_{\epsilon}^{\ast\left(n\right)}\left(NJ_{s^{\prime}}|\mathbf x^C\left(m\right)\right) \Big)$ by $\Pr\bigg(p_{Y|S}\big(\cdot|S=s\big),\mathbf \mathbb{A}_{\epsilon}^{\ast\left(n\right)}\left(NJ_{s^{\prime}}|\mathbf x^C\left(m\right)\right)\bigg)$, in \eqref{achievabilityb5} we apply Lemma \ref{lem:cstsets3} in Section \ref{subs:typicalsets} and where $M_{s^{\prime}}\in P\left(\mathbb Y\right)$ is the marginal type of $NJ_{s^{\prime}}$ and finally in \eqref{achievabilityb6} we apply the non-negativity of KL-distance,
	So, we see that as long as $n$ is large enough and $\epsilon$ small enough such that $\delta^{\prime}<\epsilon_1$ and if $R\leq \min_{s\in \mathbb S} I\left(NJ_s\right)$, $\lim_{n\to \infty}\max_{s\in\mathbb S}\Pr\left(\hat{M}\neq M|s \quad \text{is the selected channel}\right)$ $= 0$. 
	
	We continue with the interference coordination part. First, we state
	and prove the following Lemma:
	
	\begin{lemma}If $N\in \mathbf P_{pim}\left(Q,p_{Y|X}\right)$ and if the input of the DMC $p_{Y|X}$ satisfies $V\left(N,P_{\mathbf x}\right)\to 0$ in probability, then the output of the DMC satisfies $V\left(Q,P_{\mathbf y}\right)\to 0$, in probability.
	\end{lemma}
	\begin{IEEEproof}
		From Theorem \ref{th:cstsets} in Subsection \ref{subs:typicalsets}, if $\mathbf x\in 	\mathbb{A}_{\frac{\epsilon}{2|\mathbb Y|}}^{\ast\left(n\right)}\left(N\right)$, then 
		\begin{IEEEeqnarray*}{c}
			\Pr\bigg(p_{Y|X},\mathbb{A}_{\epsilon}^{\ast\left(n\right)}\left(P|\mathbf x\right)\Big|\mathbf x\bigg)>1-\delta_t\left(n,\frac{\epsilon}{2},\mathbb X\times \mathbb Y\right),	
		\end{IEEEeqnarray*} 
		where $P\triangleq NJ$ and $J\in \mathbf P^e\left(\mathbb Y|\mathbb X\right)$ such that $p_{Y|X}\doteq J$. Thus, if $V\left(N,P_{\mathbf x}\right)\to 0$ in probability, then the whole probality of $\mathbf y$ is concentrated to $\mathbb{A}_{\epsilon}^{\ast\left(n\right)}\left(P|\mathbf x\right)$. But by applying Lemma \ref{lem:cJstsets} and Lemma 	\ref{lem:Jimpliestypicality} in  Subsection \ref{subs:typicalsets} we get that $\mathbf y\in \mathbb{A}_{\epsilon}^{\ast\left(n\right)}\left(P|\mathbf x\right)\implies \left(\mathbf x,\mathbf y\right)\in \mathbb{A}_{\epsilon}^{\ast\left(n\right)}\left(P\right)\implies \mathbf y\in \mathbb{A}_{\epsilon}^{\ast\left(n\right)}\left(Q\right)$ and so the whole probability of $\mathbf y$ is also concentrated to $\mathbb{A}_{\epsilon}^{\ast\left(n\right)}\left(Q\right)$, i.e., $V\left(Q,P_{\mathbf y}\right)\to 0$ in probability.   
	\end{IEEEproof}
	Now, we will continue with our argument. We generate the codes according to $p_X\doteq N$ and thus for the input to the channel $\mathbf x$ it is true that $V\left(N,P_{\mathbf x}\right)\to 0$ in probabiltity. Due to the fact that $N\in \bigcap_{s=1}^{|\mathbb S|} \mathbf P_{\text{pim}}\left(Q_s,p_{Z|X,S}\left(\cdot|\cdot,S=s\right)\right)$ and the previous Lemma, this implies that $V\left(Q_s,P_{\mathbf z}\right)\to 0$ in probability for every $s=1,\dots,|\mathbb S|$ and as a result the second condition in the Definition \ref{def:compoundperfect1} is satisfied. %

\end{itemize}
\paragraph*{\bf Converse}
Consider a sequence of $\left(2^{nR},n\right)$-codes that achieve the rate-interference pair $\left(R,Q_1,\dots,Q_{|\mathbb S|}\right)$, i.e., according to Definition \ref{def:compoundperfect1} it satisfies \eqref{compoundperfecta} and \eqref{compoundperfectb}.
This sequence of coordination codes induces a distribution $\tilde{p}\left(z^n\right)$.
Denote as 
 $P_{\mathbf x^C\left(m\right)}$ the type of the $m$-th codeword, $m=1,\dots, 2^{nR}$ and as  $\mathbb E\left\{P_{x^n} \right\}=\mathbb E_m\left\{P_{\mathbf x^C\left(m\right)}\right\}\triangleq \sum_{m=1}^{2^{nR}} \frac{1}{2^{nR}}P_{\mathbf x^C\left(m\right)}$ the expected value of the type of the input to the channel.
We repeat the use of the scheme over $k$ blocks of length $n$ each and, as a result, the new sequence of codes consists of $ \left(2^{nR}\right)^k$ codewords $\mathbf x^{C^{\prime}}\left(m^{\prime}\right), m^{\prime}=1,\dots, \left(2^{nR}\right)^k$ of length $kn$. The new coding scheme has rate $R^\prime=\frac{\log\big( \left(2^{nR}\right)^k\big)}{kn}=R$. The type of the input to the channel which is induced by this new sequence of codes is $P_{x^{kn}}=\frac{1}{k}\sum_{i=1}^{k}P_{x^{\left(i\right)n}}\to \mathbb E\left\{P_{x^n} \right\}$ in probability.
Additionally, the encoding and the decoding process constitute of $k$ repetitions of the same process. This gives us that  $\Pr\left(\hat{M}^k\neq M^k\right)\leq k\epsilon$. So, by calling $k\epsilon\triangleq \epsilon^{\prime}$, the first condition \eqref{compoundperfecta} again is satisfied. \par The rate of this new sequence of codes should be less or equal than the mutual information $I\Big(\mathbb E\left\{P_{x^n}\right\}J_s\Big)+\epsilon_{n}$ where $\epsilon_{n}\to 0$ as $n\to \infty$, i.e., 
\begin{IEEEeqnarray}{c} R=R^\prime\leq \min_s I\Big(\mathbb E\left\{P_{x^n}\right\}J_s\Big)+\epsilon_{n}. \label{perfectconverse}\end{IEEEeqnarray}
We justify this as following:
Assume that the rate of this new sequence is for every $n$ greater or equal than $\min_s I\Big(\mathbb E\left\{P_{x^n}\right\}J_s\Big)+\epsilon$ for some $\epsilon>0$. Every codeword $x^{C^{\prime}}\left(m^{\prime}\right)$ is mapped due to Theorem \ref{th:cstsets3} in Section \ref{subs:typicalsets}  (with probability which tends to 1) to 
more than $\bigg(1-\delta_t\left(n,\frac{\epsilon}{2},\mathbb X\times \mathbb Y\right)\bigg)e^{n\big(H_{\mathbb E\left\{P_{x^n}\right\}}\left(J_s\right)-\epsilon_m\left(p^s_{X,Y}\right)\big)}$ sequences $\mathbf y$ where $p_{X,Y}\in \mathbf P\left(\mathbb X, \mathbb Y\right)$ is a joint PMF such that $p^s_{X,Y}\doteq E\left\{P_{x^n}\right\}J_s $. So for all $s$, all the codewords together are mapped in total to more than $e^{n\Big( I\big(\mathbb E\left\{P_{x^n}\right\}J_s\big)+\epsilon\Big)}\bigg(1-\delta_t\left(n,\frac{\epsilon}{2},\mathbb X\times \mathbb Y\right)\bigg)e^{n\big(H_{\mathbb E\left\{P_{x^n}\right\}}\left(J_s\right)-\epsilon_m\left(p^s_{X,Y}\right)\big)}=\bigg(1-\delta_t\left(n,\frac{\epsilon}{2},\mathbb X\times \mathbb Y\right)\bigg)e^{n\big(H\left(M_s\right)+\epsilon-\epsilon_m\left(p^s_{X,Y}\right)\big)}$ sequences $\mathbf y$ with probability which tends to 1. By $M_s\in \mathbf P^e\left(\mathbb Y\right)$ we denote here the marginal type of $\mathbb E\left\{P_{x^n}\right\}J_s$. For every such sequence $\mathbf y$, it is true (with probability which tends to 1 due to Theorem \ref{th:cstsets} in  Section \ref{subs:typicalsets}) that $\mathbf y\in \mathbb{A}_{\epsilon}^{\ast\left(n\right)}\Big(\mathbb E\left\{P_{x^n}\right\}J_s|\mathbf x\Big)\implies \left(\mathbf x,\mathbf y\right)\in \mathbb{A}_{\epsilon}^{\ast\left(n\right)}\Big(\mathbb E\left\{P_{x^n}\right\}J_s\Big)\implies \mathbf y\in \mathbb{A}_{\epsilon}^{\ast\left(n\right)}\left(M_s\right)$ due to Lemma \ref{lem:cJstsets} and Lemma 	\ref{lem:Jimpliestypicality} in  Section \ref{subs:typicalsets}. Therefore, all the sequences which can be introduced by a codeword $x^{C^{\prime}}\left(m^{\prime}\right)$ in the space $\mathbb Y$ are less than $e^{n\big(H\left(M_s\right)+\epsilon_m\left(p^s_{Y}\right)\big)}$
 due to Theorem \ref{th:Jstsets3} in  Section \ref{subs:typicalsets}. Here, $p^s_{Y}\in \mathbf P\left(\mathbb Y\right)$ is a PMF such that $p^s_{Y}\doteq M_s$. We conclude that more than $e^{nH\left(M_s\right)}\left(e^{n\epsilon}-1\right)$ sequences can be introduced by more than one codewords and thus they cannot be decoded uniquely (we omit some asymptotically unimportan terms). Each of these sequences has probability greater or equal than $\frac{1}{M}e^{-n\big(H_{\mathbb E\left\{P_{x^n}\right\}}\left(J_s\right)+\epsilon_m\left(p^s_{X,Y}\right)\big)}$ $\geq\frac{1}{e^{n\Big( I\big(\mathbb E\left\{P_{x^n}\right\}J_s\big)+\epsilon\Big)}}e^{-n\big(H_{\mathbb E\left\{P_{x^n}\right\}}\left(J_s\right)+\epsilon_m\left(p^s_{X,Y}\right)\big)}$ due to Theorem \ref{th:cstsets1} in  Section \ref{subs:typicalsets}  and so this overall error probability is greater than $\frac{1}{e^{n\Big( I\big(\mathbb E\left\{P_{x^n}\right\}J_s\big)+\epsilon\Big)}}e^{-n\big(H_{\mathbb E\left\{P_{x^n}\right\}}\left(J_s\right)+\epsilon_m\left(p^s_{X,Y}\right)\big )}\cdot \cdot e^{nH\left(M_s\right)}\left(e^{n\epsilon}-1\right)$ which tends to one and therefore we have reached in a contradiction. \par Subsequently, we state without proof the following Lemma which is a direct generalization of \cite[Lemma 10]{serrano:2014}.
\begin{lemma}
	Let $Q_1,\dots,Q_{|\mathbb S|}\in P^e\left(\mathbb Z\right)$ fixed and each have pre-image $\mathbf P_{\text{pim}}\left(Q_s,p_{Z|X,S}\left(\cdot|\cdot,S=s\right)\right)$ respectively. If a sequence of $\left(2^{nR},n\right)$-codes induces an interference type $T_{z^n}$ such that $V\left(Q_s,T_{z^n}\right)\to 0, \text{ in probability if} \quad S=s, \forall s\in \mathbb S$,
	then the expectation of the type of the codewords $\mathbb E \left\{P_{x^n}\right\}$ satisfies
	$V\Big(\mathbb E \left\{P_{x^n}\right\},P_X^{\left(n\right)}\Big)\to 0$, for some sequence $P_X^{\left(n\right)}$ with $P_X^{\left(n\right)}\in \bigcap_{s=1}^{|\mathbb S|} \mathbf P_{\text{pim}}\left(Q_s,p_{Z|X,S}\left(\cdot|\cdot,S=s\right)\right)$ for all $n$.
\end{lemma}
\par By applying the previous Lemma to the sequence of codes $C$, we conclude that there exists a sequence $P_X^{\left(n\right)}\in \bigcap_{s=1}^{|\mathbb S|} \mathbf P_{\text{pim}}\left(Q_s,p_{Z|X,S}\left(\cdot|\cdot,S=s\right)\right)$ for all $n$ such that $V\left(\mathbb E \left\{P_{x^n}\right\}, P_X^{\left(n\right)}\right)\to 0$ and so 
$V\left(\mathbb E \left\{P_{x^n}\right\}J_s, P_X^{\left(n\right)}J_s\right)\to 0$
for all $s=1,\dots, |\mathbb S|$. Since mutual information is a continuous function of the input type, this convergence together with \eqref{perfectconverse} implies that 

\begin{IEEEeqnarray*}{rCl}
	R&\leq& \limsup_{n\to \infty}\min_sI\Big(\mathbb E\left\{P_{x^n}\right\}J_s\Big)=\limsup_{n\to \infty}\min_sI\left(P_X^{\left(n\right)}J_s\right)\\&\leq&\max_{N\in \bigcap_{s=1}^{|\mathbb S|} \mathbf P_{\text{pim}}\left(Q_s,p_{Z|X,S}\left(\cdot|\cdot,S=s\right)\right)} \min_sI\left(NJ_s\right). 
\end{IEEEeqnarray*}
This completes the proof.
\section{Proof of Theorem \ref{th:compoundperfectimperfect}}
\label{subs:proofs3}
\paragraph*{\bf Achievability} 
When a rate-interference tuple $\left(R,N_1,\dots,N_{|\mathbb S|}\right)$ is in the interior of $C$ for some $\left(N_1,\dots,N_{|\mathbb S|}\right)$ such that $ \left(N_1,\dots,N_{|\mathbb S|}\right) \in  N_{\Delta_1}\big(Q\big)\times \dots\times N_{\Delta_{|\mathbb S|}}\big(Q\big) $, we are assured (see Definition \ref{def:compoundperfect1})
the existence of a  communication-coordination code such  that for every $\epsilon_0,\epsilon_1,\dots,\epsilon_{|\mathbb S|}>0$ and for all large enough $n$ we have
\begin{IEEEeqnarray*}{rCl}\Pr\left(\hat{M}\neq M\right)&<& \epsilon_0,\\\Pr\Big(V\left(N_s,P_{\mathbf z}\right)> \epsilon_s\Big)&<&\epsilon_s, \quad \forall s=1,\dots,|\mathbb S|.
\end{IEEEeqnarray*}
Applying the triangle inequality on the variational distance and because  $N_s\in  N_{\Delta_s}\big(Q\big)$, we obtain
\begin{IEEEeqnarray*}{rCl}V\left(Q,P_{\mathbf z}\right)&\leq& V\left(N_s,P_{\mathbf z}\right)+V\left(N_s,Q\right)\\&\leq&V\left(N_s,P_{\mathbf z}\right) +\Delta_s\nonumber.	\end{IEEEeqnarray*}
Thus, for every $\epsilon_s>0,\quad s=1,\dots,|\mathbb S|$ and for all large enough $n$, this communication-coordination code achieves
$	\Pr\Big\{V\left(Q,P_{\mathbf z}\right)>\Delta_s+\epsilon_s\Big\}<\epsilon_s,$
which gives 
\begin{multline*}
	\mathbb{E}_{\tilde p^s_{Z^n}}\Big\{V\left(Q,P_{\mathbf z}\right)\Big\}\leq\Pr\Big\{V\left(Q,P_{\mathbf z}\right)>\Delta_s+\epsilon_s\Big\}\times V_{\max}\\\quad \quad \quad \quad \quad \quad \quad \quad  +\Pr\Big\{V\left(Q,P_{\mathbf z}\right)\leq\Delta_s+\epsilon_s\Big\}\times \left(\Delta_s+\epsilon_s\right)\\\leq \Delta_s+\epsilon_s+\epsilon_s V_{\max}.
\end{multline*}
By choosing $\epsilon_s$ arbitrarily small and $n$ large enough, we conclude that this communication-coordination code achieves also $\Delta_1\cdots \Delta_{|\mathbb S|}$-interference coordination according to $Q$ and so $\left(R,\Delta_1,\dots, \Delta_{|\mathbb S|}\right)\in R^I\big(Q\big)$.\paragraph*{\bf Converse}
Suppose that $\left(R,\Delta_1,\dots,\Delta_{|S|}\right)$ is in the interior of  $R^I\big(Q\big)$, i.e., for all $\epsilon>0$, there exists an $N$ such that for all $n>N$, there exists  a communication-coordination code with blocklength $n$ such that
\begin{IEEEeqnarray*}{rCl}&& \max_{s\in\mathbb S}\Pr\left(\hat{M}\neq M|s \quad \text{is the selected channel}\right)< \epsilon,\\&& \mathbb E_{\tilde p^s_{Z^n}}\left\{V\left(Q,P_{z^n}\right)\right\}\leq \Delta_s,\quad \forall S=s.
\end{IEEEeqnarray*}
This  communication-coordination code induces a tuple of distributions $\left(\tilde{p}^1_{Z^n},\dots,\tilde{p}^{|\mathbb S|}_{Z^n}\right)$.
We repeat the use of the scheme over $k$ blocks of length $n$ each and, as a result, we induce a sequence of joint distributions on $Z^{kn}$ and each consists of blocks $\left(Z_1^{n},\dots,Z_{kn-n+1}^{kn}\right)$.
The new coding scheme has rate $R^\prime=\frac{\log\big(\left(2^{nR}\right)^k\big)}{kn}=R$.
So, the encoding and the decoding process constitute of $k$ repetitions of the same process. This gives us that  $\Pr\left(\hat{M}^k\neq M^k\right)\leq k\epsilon$. So, by calling $k\epsilon\triangleq \epsilon^{\prime}$, the first condition of Definition \ref{def:compoundperfect1} is satisfied.
Additionally, by the law of large numbers, we get $P_{z^{kn}}=\frac{1}{k}\sum_{i=1}^{k}{P_{z^{\left(i\right)n}}}\to \mathbb E_{\tilde p^s_{Z^n}}\left\{P_{z^n}\right\},$
in probability for every $s=1,\dots,|\mathbb S|$.
Point-wise convergence in probability further implies that as $k$ grows
$	V\big(P^s_{z^{kn}},\mathbb E_{\tilde p^s_{Z^n}}\left\{P_{z^n}\right\}\big)\to 0$
in probability.
However,
$	\mathbb E_{\tilde p^s_{Z^n}}\left\{P_{z^n}\right\}=N_s$,
and thus
$V\left(P_{z^{kn}},N_s\right)\to 0$,
in probability,
where $N_s\in P^e\left(\mathbb Z\right)$ is a type such that $\frac{1}{n}\sum_{k=1}^n\tilde p^s_{Z_k}\doteq N_s$
(see  Lemma \ref{lem:mainlem} in Section \ref{subs:converselemma}). 
Moreover $\Delta_s\geq\mathbb{E}_{\tilde p^s_{Z^n}}\big\{V\left(Q,P_{ z^n}\right)\big\}\stackrel{(a)}{\geq} V\Big(Q,\mathbb{E}_{\tilde p^s_{Z^n}}\left\{P_{z^n}\right\}\Big) 
=V\left(Q,N_s\right),$
where ($a$) follows from Jensen's inequality since variational distance is convex, i.e., for every $Q,S,N\in P^e\left(\mathbb Z\right)$ and for every $\lambda\in \left[0,1\right]$ we have
\begin{IEEEeqnarray*}{c}\lambda V\left(Q,S\right)+\left(1-\lambda\right)V\left(Q,N\right)\geq V\Big(Q,\big(\lambda S+\left(1-\lambda\right)N\big)\Big).\end{IEEEeqnarray*}
Thus, we have constructed a sequence of communication-coordination codes with rate $R$ that achieves multiple interference coordination according to $\left(N_1,\dots,N_s\right)$, i.e., $\left(R,N_1,\dots,N_{|\mathbb S|}\right)\in C $ and additionally
$\left(N_1,\dots,N_{|\mathbb S|}\right)\in N_{\Delta_1}\left(Q\right)\times\dots\times N_{\Delta_{|\mathbb S|}}\left(Q\right)$. This completes the proof.
\section{Additional Mathematical Background}
\label{subs:proofs1}
\subsection{Types, PMFs and chain rules}

\par Joint types and types satisfy the following:
$P_{\mathbf x}\left(a\right)=\sum_{b\in \mathbb Y}P_{\mathbf x,\mathbf y}\left(a,b\right) \quad \forall  a\in\mathbb X$ and $P_{\mathbf y}\left(b\right)=\sum_{a\in \mathbb Y}P_{\mathbf x,\mathbf y}\left(a,b\right) \quad \forall  b\in\mathbb Y$.
As a consequence, for a given joint type $P\in \mathbf P^e\left(\mathbb X,\mathbb Y\right)$, we define its marginal types $Q\in \mathbf P^e\left(\mathbb X\right)$ and $N\in \mathbf P^e\left(\mathbb Y\right)$ with $Q\left(a\right)\triangleq \sum_{b\in \mathbb Y}P\left(a,b\right), \forall a\in \mathbb X$ and $N\left(b\right)\triangleq \sum_{a\in \mathbb Y}P\left(a,b\right), \forall b\in \mathbb Y$. Analogously, for a joint PMF  $p_{X,Y}\in \mathbf P\left(\mathbb X, \mathbb Y\right)$, we define its marginal PMFs $p_X\in \mathbf P\left(\mathbb X\right)$ with $p_X\left(a\right)\triangleq \sum_{b\in \mathbb Y}p_{X,Y}\left(a,b\right), \forall a\in \mathbb X$ and $p_Y\in \mathbf P\left(\mathbb Y\right)$ with $p_Y\left(b\right)\triangleq \sum_{a\in \mathbb Y}p_{X,Y}\left(a,b\right), \forall b\in \mathbb Y$. 
\par Conditional types, joint types and types satisfy the following chain rules: $P_{\mathbf y|\mathbf x}\left(b|a\right)= \frac{N_o\left(a,b|\mathbf x,\mathbf y\right)}{N_o\left(a|\mathbf x\right)}=\frac{\frac{N_o\left(a,b|\mathbf x,\mathbf y\right)}{n}}{\frac{N_o\left(a|\mathbf x\right)}{n}}=\frac{P_{\mathbf x,\mathbf y}\left(a,b\right)}{P_{\mathbf x}\left(a\right)}$ and $P_{\mathbf x|\mathbf y}\left(a|b\right)= \frac{N_o\left(a,b|\mathbf x,\mathbf y\right)}{N_o\left(b|\mathbf y\right)}=\frac{\frac{N_o\left(a,b|\mathbf x,\mathbf y\right)}{n}}{\frac{N_o\left(b|\mathbf y\right)}{n}}=\frac{P_{\mathbf x,\mathbf y}\left(a,b\right)}{P_{\mathbf y}\left(b\right)}$.
As a consequence, for a given joint type $P\in \mathbf P^e\left(\mathbb X,\mathbb Y\right)$, we define its marginal conditional types $J\in \mathbf P^e\left(\mathbb Y|\mathbb X\right)$ with $J\left(b|a\right)=\frac{P\left(a,b\right)}{Q\left(a\right)}$ and $T\in \mathbf P^e\left(\mathbb X|\mathbb Y\right)$ with $T\left(a|b\right)=\frac{P\left(a,b\right)}{N\left(b\right)}$, $ \forall a\in \mathbb X, b\in \mathbb Y,$ where $Q \in \mathbf P^e\left(\mathbb X\right)$ and $N\in \mathbf P^e\left(\mathbb Y\right)$ are the marginal types of $P$. Analogously, for a joint PMF  $p_{X,Y}\in \mathbf P\left(\mathbb X, \mathbb Y\right)$, we define its marginal conditional PMFs $p_{Y|X}\in \mathbf P\left(\mathbb Y|\mathbb X\right)$ with  $p_{Y|X}\left(b|a\right)=\frac{p_{X,Y}\left(a,b\right)}{P_X\left(a\right)}$ and $p_{X|Y}\left(x|y\right)\in \mathbf P\left(\mathbb X|\mathbb Y\right)$ with  $P_{X|Y}\left(a|b\right)=\frac{p_{X,Y}\left(a,b\right)}{P_Y\left(b\right)}$, $ \forall \left(a,b\right) \in \mathbb X \times \mathbb Y$. 
Conversely, for a given type $Q\in \mathbf P^e\left(\mathbb X\right)$ and a given conditional type $J\in \mathbf P^e\left(\mathbb Y|\mathbb X\right)$ we define the joint type $P\triangleq QJ$ with $P\left(a,b\right)\triangleq Q\left(a\right)J\left(b|a\right), \forall \left(a,b\right)\in \mathbb X\times \mathbb Y$. Analogously, for a given PMF $p_X\in \mathbf P\left(\mathbb X\right)$ and a given conditional PMF $p_{Y|X}\in \mathbf P\left(\mathbb Y|\mathbb X\right)$ we define the joint PMF $p_{X,Y}\triangleq p_Xp_{Y|X}$ with $p_{X,Y}\left(a,b\right)=p_X\left(a\right)p_{Y|X}\left(b|a\right), \forall \left(a,b\right)\in \mathbb X\times \mathbb Y$.

\par We adopt the following notation:
$\Pr\left(p_X,\mathbf x\right)\triangleq \prod_{i=1}^{n}p_X\left(x_i\right), \quad \Pr\left(p_X,\mathbf A\right)\triangleq\sum_{\mathbf x\in \mathbf A} \Pr\left(p_X,\mathbf x\right) \quad \text{for}\quad \mathbf  A \subseteq \mathbb X^n, \quad \Pr\bigg(p_{Y|X},\mathbf y\Big|\mathbf x\bigg)\triangleq \prod_{i=1}^{n}p_{Y|X}\left(y_i|x_i\right),\quad \Pr\bigg(p_{Y|X}, \mathbf A\Big|\mathbf x\bigg)\triangleq\sum_{\mathbf y\in \mathbf A}\Pr\bigg(p_{Y|X},\mathbf y\Big|\mathbf x\bigg) \quad \text{for}\quad \mathbf A \subseteq \mathbb Y^n.$

\subsection{Entropy, Conditional Entropy, KL-Distance and Mutual Information}

\begin{definition}[Joint and conditional entropy] 
	Let a type $Q\in \mathbf P^e\left(\mathbb X\right)$, a conditional type $J\in \mathbf P^e\left(\mathbb Y|\mathbb X\right)$  and a joint type $P\in \mathbf P^e\left(\mathbb X,\mathbb Y\right) $. We define the joint entropy of $P$ and the conditional entropy of $J$ with respect to $Q$ respectively:
	\begin{IEEEeqnarray*}{rCl}H\left(P\right)&\triangleq&\sum_{\left(a,b\right)\in \mathbb X\times \mathbb Y} P\left(a,b\right)\log \frac{1}{P\left(a,b\right)}.\label{jentropy}\\H_Q\left(J\right)&\triangleq&\sum_{a\in \mathbb X} Q\left(a\right)H\big(J\left(\cdot|a\right)\big).\label{centropy} 
	\end{IEEEeqnarray*}
\end{definition}
\begin{definition}[KL-divergence] Let a joint PMF $p_{X,Y}\in \mathbf P\left(\mathbb X,\mathbb Y\right) $ and a joint type $P\in \mathbf P^e\left(\mathbb X,\mathbb Y\right) $. We define the KL-divergence between $p_{X,Y}$ and $P$ as:
	\begin{IEEEeqnarray*}{c} D\left(P||p_{X,Y}\right)\triangleq\sum_{\left(a,b\right)\in \mathbb X\times \mathbb Y} P\left(a,b\right)\log \frac{P\left(a,b\right)}{B\left(a,b\right)},\label{KLdistance} \end{IEEEeqnarray*}
	where $B\in \mathbf P^e\left(\mathbb X,\mathbb Y\right)$ is a joint type such that $p_{X,Y}\doteq B$.
\end{definition}

\begin{definition}[Mutual Information]
	Let a  joint type $P\in \mathbf P^e\left(\mathbb X,\mathbb Y\right)$. We define the mutual information of $P$ as follows:
	\begin{IEEEeqnarray*}{c} I\left(P\right)\triangleq D\left(P||p_{X}p_{Y}\right),\label{minformation}
	\end{IEEEeqnarray*}
	where $p_{X,Y}\in \mathbf P\left(\mathbb X,\mathbb Y\right)$ is a joint PMF such that $p_{X,Y}\doteq P$ and $p_X \in \mathbf P\left(\mathbb X\right)$ and $p_Y \in \mathbf P\left(\mathbb Y\right)$ are the marginal distributions of $p_{X,Y}$.
\end{definition}
\begin{remark}[Mutual information, entropy and conditional entropy]
	Let a joint type $P\in \mathbf P^e\left(\mathbb X, \mathbb Y\right)$, its marginal types $Q\in \mathbf P^e\left(\mathbb X\right)$, $N\in \mathbf P^e\left(\mathbb Y\right)$ and its marginal conditional types $J\in \mathbf P^e\left(\mathbb Y|\mathbb X\right)$, $T\in \mathbf P^e\left(\mathbb X|\mathbb Y\right)$ . Then, we can prove that 
	\begin{IEEEeqnarray*}{c} I\left(P\right)=H\left(Q\right)-H_N\left(T\right)=H\left(N\right)-H_Q\left(J\right). \label {minformation1}\end{IEEEeqnarray*}
\end{remark}

\section{A useful Lemma}
\label{subs:converselemma}
\begin{lemma}Let a $n$-length sequence $\mathbf x$ which is generated according to a PMF $p_{X^n}\in \mathbf P\left(\mathbb X^n\right)$. 
	Then,
	\begin{IEEEeqnarray*}{c}\mathbb{E}_{p_{X^n}}\big\{P_{\mathbf x}\big\}= N,
	\end{IEEEeqnarray*}
	where $N\in \mathbf P^e\left(\mathbb X\right)$ is a type such that \begin{IEEEeqnarray*}{rCl}\frac{1}{n}\Bigg(\sum_{k=1}^n p_{X_k}\Bigg)\doteq N.\end{IEEEeqnarray*}
	\label{lem:mainlem}	
\end{lemma}
\begin{IEEEproof}
	For all $a\in \mathbb X$, we get
	\begin{IEEEeqnarray*}{rCl}
		\mathbb{E}_{p_{X^n}}\big\{P_{\mathbf x}\big\}\left(a\right)&=&\sum_{\mathbf x}p_{X^n}\left(\mathbf x\right)\frac{N_o\left(a|\mathbf x\right)}{n} \\
		&=&\sum_{\mathbf x}\Bigg(p_{X^n}\left(\mathbf x\right)\frac{1}{n}\sum_{k=1}^n\mathbf{1}\left(x_k=a\right)\Bigg)\\
		&=&\sum_{k=1}^n\sum_{\mathbf x}\Bigg(p_{X^n}\left(\mathbf x\right)\frac{1}{n}\mathbf{1}\left(x_k=a\right)\Bigg)\\
		&=&\frac{1}{n}\sum_{k=1}^n\Big(p_{X_k}\left(a\right)\Big)\\&=&\frac{1}{n}\Bigg(\sum_{k=1}^n p_{X_k}\Bigg)\left(a\right)\\&=&N\left(a\right).	\end{IEEEeqnarray*}
\end{IEEEproof}
\section{Typical Sets}
\label{subs:typicalsets}
The material of this section has been taken mainly from \cite [Chapter 4]{moser:2021}. The notation is quite different and is based on the definitions that we developed in Section \ref{s:preliminaries} and Appendix \ref{subs:proofs1}.
\begin{definition}[Strongly jointly $\epsilon$-typical set]
	Fix an $\epsilon>0$, an $n$ and a joint type $P\in \mathbf P^e\left(\mathbb X,\mathbb Y\right)$. The strongly jointly $\epsilon$-typical set $\mathbb{A}_{\epsilon}^{\ast\left(n\right)}\left(P\right)$ with respect to the joint type $P$ is defined as: 
	
	\begin{multline*}
	\mathbb{A}_{\epsilon}^{\ast\left(n\right)}\left(P\right)\triangleq\\
	\left.
	\left\{ \,
	\begin{IEEEeqnarraybox}[
	\IEEEeqnarraystrutmode
	\IEEEeqnarraystrutsizeadd{2pt}
	{2pt}][c]{l}
	\left(\mathbf x,\mathbf y\right) \in \mathbb X^n\times \mathbb Y^n:\\
	|P_{\mathbf x,\mathbf y}\left(a,b\right)-P\left(a,b\right)|
	<\frac{\epsilon}{|\mathbb X||\mathbb Y|},\forall \left(a,b\right)\in \mathbb X\times \mathbb Y, \text{and}\\ P_{\mathbf x,\mathbf y}\left(a,b\right)=0 , \quad \forall \left(a,b\right)\in \mathbb X\times \mathbb Y \quad \text{with} \quad P\left(a,b\right)=0
	\end{IEEEeqnarraybox}\right\}.
	\right.
	\label{def:Jstsets}
	\end{multline*}	
\end{definition}
\begin{lemma} Let a joint type $P\in \mathbf P^e\left(\mathbb X, \mathbb Y\right)$ and its marginal types $Q\in \mathbf P^e\left(\mathbb X\right)$ and $S\in \mathbf P^e\left(\mathbb Y\right)$. Then,  
	\begin{IEEEeqnarray*}{c}
		\left(\mathbf x,\mathbf y\right)\in 	\mathbb{A}_{\epsilon}^{\ast\left(n\right)}\left(P\right)\implies \mathbf x\in \mathbb{A}_{\epsilon}^{\ast\left(n\right)}\left(Q\right) \quad \text{and} \quad  \mathbf y\in \mathbb{A}_{\epsilon}^{\ast\left(n\right)}\left(S\right).
	\end{IEEEeqnarray*}
	\label{lem:Jimpliestypicality}
\end{lemma}
\begin{definition}[$\epsilon_m,\delta_t$]
	Let a joint  PMF $p_{X,Y}\left(x,y\right)\in  \mathbf P\left(\mathbb X,\mathbb Y\right)$.
	We name one particular $\epsilon$ and one particular $\delta$ that we meet often:
	\begin{IEEEeqnarray*}{rCl}&&\epsilon_m\left(p_{X,Y}\right)\triangleq-\epsilon\log\left(p_{X,Y}^{\min}\right),\label{Jem}\\&&\delta_{t}\left(n,\epsilon,\mathbb{X}\times \mathbb Y\right)\triangleq \left(n+1\right)^{|\mathbb X||\mathbb Y|}e^{-n\frac{\epsilon^2}{2|\mathbb X|^2|\mathbb Y|^2}\log e},\label{Jdeltat}
	\end{IEEEeqnarray*}
	where $p_{X,Y}^{\min}$ denotes the smallest positive value of $p_{X,Y}\left(a,b\right)$.
	\label{def:epsilondelta}
\end{definition}
\begin{theorem}[Bounding the size of a strongly jointly typical set]Let $P\in \mathbf P^e\left(\mathbb X, \mathbb Y\right)$. The size of the strongly jointly typical set $\mathbb{A}_{\epsilon}^{\ast\left(n\right)}\left(P\right)$ is bounded as follows:
	\begin{multline*}\big(1-\delta_t\left(n,\epsilon,\mathbb X\times \mathbb Y\right)\big)e^{n\big(H\left(P\right)-\epsilon_m\left(p_{X,Y}\right)\big)}\leq  |\mathbb{A}_{\epsilon}^{\ast\left(n\right)}\left(P\right)|\\  \leq e^{n\big(H\left(P\right)+\epsilon_m\left(p_{X,Y}\right)\big)},
	\end{multline*}
	where $p_{X,Y}\in \mathbf P\left(\mathbb X,\mathbb Y\right)$ is a joint PMF such that $p_{X,Y}\doteq P$.
	\label{th:Jstsets3}
\end{theorem}
\begin{definition}[Strongly conditionally typical sets with respect to a joint type]
	Let a joint type $P\in \mathbf P^e\left(\mathbb X, \mathbb Y\right)$, its marginal type $Q\in \mathbf P^e\left(\mathbb X\right)$ and a typical sequence $\mathbf x\in \mathbb{A}_{\epsilon}^{\ast\left(n\right)}\left(Q\right).$ We define the strongly conditionally typical set with respect to $P$ and $\mathbf x$ as
	\begin{IEEEeqnarray*}{c}\mathbb{A}_{\epsilon}^{\ast\left(n\right)}\left(P|\mathbf x\right)\triangleq\left\{\mathbf y\in \mathbb Y^n:\left(\mathbf x, \mathbf y\right)\in \mathbb{A}_{\epsilon}^{\ast\left(n\right)}\left(P\right) \right\}.\label{def:cstsets}
	\end{IEEEeqnarray*}
\end{definition}
\begin{lemma}Let a joint type $P\in \mathbf P^e\left(\mathbb X, \mathbb Y\right)$ and its marginal type $Q\in \mathbf P^e\left(\mathbb X\right)$.
	The event $\Big\{\left(\mathbf x,\mathbf y\right)\in \mathbb{A}_{\epsilon}^{\ast\left(n\right)}\left(P\right) \Big\}$
	is equivalent to the event $\left\{\mathbf x\in \mathbb{A}_{\epsilon}^{\ast\left(n\right)}\left(Q\right) \right\}\cap \left\{\mathbf y\in \mathbb{A}_{\epsilon}^{\ast\left(n\right)}\big(P|\mathbf x\big)\right\}$.
	\label{lem:cJstsets}
\end{lemma}
\begin{theorem}[Probability of a strongly conditionally typical sequence]
	Let a joint PMF $p_{X,Y}\in \mathbf P\left(\mathbb X, \mathbb Y\right)$ and a joint type $P\in  \mathbf P^e\left(\mathbb X,\mathbb Y\right)$, such that $p_{X,Y}\doteq P$. Let also the marginal conditional PMF $p_{Y|X}\in \mathbf P\left(\mathbb Y|\mathbb X\right)$  of $p_{X,Y}$, the marginal type $Q\in \mathbf P^e\left(\mathbb X\right)$ and the marginal conditional type $J\in \mathbf P^e\left(\mathbb Y|\mathbb X\right)$ of $P$. Also consider a typical sequence $\mathbf x\in 	\mathbb{A}_{\epsilon}^{\ast\left(n\right)}\left(Q\right) $. The probability of $\mathbf y\in \mathbb{A}_{\epsilon}^{\ast\left(n\right)}\left(P|\mathbf x\right) $ given $\mathbf x$ under $p_{Y|X}$ is bounded as follows:
	\begin{IEEEeqnarray*}{rCl}e^{-n\big(H_Q\left(J\right)+\epsilon_m\left(p_{X,Y}\right)\big)}&\leq& \Pr\bigg(p_{Y|X},\mathbf y\Big|\mathbf x\bigg)\\&\leq& e^{-n\big(H_Q\left(J\right)-\epsilon_m\left(p_{X,Y}\right)\big)}. \label{cpsequence}
	\end{IEEEeqnarray*}
	\label{th:cstsets1}
\end{theorem}
\begin{theorem}[Total probability of a conditionally strongly typical set]
	
	Let a joint PMF $p_{X,Y}\in \mathbf P\left(\mathbb X, \mathbb Y\right)$ and a joint type $P\in  \mathbf P^e\left(\mathbb X,\mathbb Y\right)$ such that $p_{X,Y}\doteq P$. Let also the marginal conditional PMF $p_{Y|X}\in \mathbf P\left(\mathbb Y|\mathbb X\right)$ of $p_{X,Y}$, the marginal type $Q\in \mathbf P^e\left(\mathbb X\right)$ and the marginal conditional type $J\in \mathbf P^e\left(\mathbb Y|\mathbb X\right)$ of $P$.  Consider also a typical sequence $\mathbf x\in 	\mathbb{A}_{\epsilon}^{\ast\left(n\right)}\left(Q\right)$. Then, the total probability of the strongly conditionally typical set $\mathbb{A}_{\epsilon}^{\ast\left(n\right)}\left(P|\mathbf x\right)$  under $p_{Y|X}$ is bounded trivially as:
	\begin{IEEEeqnarray*}{c}
		\Pr\bigg(p_{Y|X},\mathbb{A}_{\epsilon}^{\ast\left(n\right)}\left(P|\mathbf x\right)\Big|\mathbf x\bigg)\leq 1. \label{ctotprob51}
	\end{IEEEeqnarray*}
	If in addition $\mathbf x\in 	\mathbb{A}_{\frac{\epsilon}{2|\mathbb Y|}}^{\ast\left(n\right)}\left(Q\right)$, then we can also provide the lower bound 
	\begin{IEEEeqnarray*}{c}
		\Pr\bigg(p_{Y|X},\mathbb{A}_{\epsilon}^{\ast\left(n\right)}\left(P|\mathbf x\right)\Big|\mathbf x\bigg)>1-\delta_t\left(n,\frac{\epsilon}{2},\mathbb X\times \mathbb Y\right).	 \label{ctotprob52}
	\end{IEEEeqnarray*}
	\label{th:cstsets}
\end{theorem}

\begin{theorem}[Size of a strongly conditionally typical set] Let $P\in \mathbf P^e\left(\mathbb X, \mathbb Y\right)$, its marginal type $Q\in \mathbf P^e\left(\mathbb X\right)$, its marginal conditional type $J\in \mathbf P^e\left(\mathbb Y|\mathbb X\right)$  and a given sequence  $\mathbf x\in \mathbb{A}_{\epsilon}^{\ast\left(n\right)}\left(Q\right) $. The size of the strongly conditionally typical set $\mathbb{A}_{\epsilon}^{\ast\left(n\right)}\left(P|\mathbf x\right)$ is bounded as follows:
	\begin{IEEEeqnarray*}{rCl} |\mathbb{A}_{\epsilon}^{\ast\left(n\right)}\left(P|\mathbf x\right)|  \leq e^{n\big(H_Q\left(J\right)+\epsilon_m\left(p_{X,Y}\right)\big)}, \label{ccardinality}
	\end{IEEEeqnarray*}
	where $p_{X,Y}\in \mathbf P\left(\mathbb X,\mathbb Y\right)$ is a joint PMF such that $p_{X,Y}\doteq P$. 
	If in addition $\mathbf x\in \mathbb{A}_{\frac{\epsilon}{2|\mathbb Y|}}^{\ast\left(n\right)}\left(Q\right)$, then we can also provide the lower bound:
	\begin{IEEEeqnarray*}{rCl} &&|\mathbb{A}_{\epsilon}^{\ast\left(n\right)}\left(P|\mathbf x\right)|\nonumber\\&&>\bigg(1-\delta_t\left(n,\frac{\epsilon}{2},\mathbb X\times \mathbb Y\right)\bigg)e^{n\big(H_Q\left(J\right)-\epsilon_m\left(p_{X,Y}\right)\big)}\label{ccardinalitye}.
	\end{IEEEeqnarray*}
	\label{th:cstsets3}
\end{theorem}
\begin{lemma}[Total probability of $\mathbb{A}_{\epsilon}^{\ast\left(n\right)}\left(P|\mathbf x\right)$ under $q_{Y}$]
	Let a joint type $P\in \mathbf P^e\left(\mathbb X,\mathbb Y\right)$, its marginal types $Q\in  \mathbf P^e\left(\mathbb X\right)$, $S\in  \mathbf P^e\left(\mathbb Y\right)$ 
	and a sequence $\mathbf x \in \mathbb{A}_{\epsilon}^{\ast\left(n\right)}\left(Q\right)$.
	Consider also the PMF 
	$q_Y\in \mathbf P\left(\mathbb Y\right)$. Then,
	\begin{IEEEeqnarray*}{c}
		\Pr\big(q_Y,\mathbb{A}_{\epsilon}^{\ast\left(n\right)}\left(P|\mathbf x\right)\big)<e^{-n\big(I\left(P\right)+D\left(S||q_Y\right)-\epsilon_m\left(p_{X,Y}\right)-\epsilon_m\left(q_Y\right)\big)},\label{cconsequences2}
	\end{IEEEeqnarray*}
	where $p_{X,Y}\in \mathbf P\left(\mathbb X,\mathbb Y\right)$ is a joint PMF such that $p_{X,Y}\doteq P$.
	If in addition also  $\mathbf x\in \mathbb{A}_{\frac{\epsilon}{2|\mathbb Y|}}^{\ast\left(n\right)}\left(Q\right)$, then we can also provide the lower bound:
	\begin{IEEEeqnarray*}{rCl}&&\Pr\big(q_Y,\mathbb{A}_{\epsilon}^{\ast\left(n\right)}\left(P|\mathbf x\right)\big)>\bigg(1-\delta_t\left(n,\frac{\epsilon}{2},\mathbb X\times \mathbb Y\right)\bigg)\\&&\quad \quad \quad \quad \quad \quad \quad \quad \quad  e^{-n\big(I\left(P\right)+D\left(S||q_Y\right)+\epsilon_m\left(p_{X,Y}\right)+\epsilon_m\left(q_Y\right)\big)}. \label{cconsequences3}
	\end{IEEEeqnarray*}
	\label{lem:cstsets3}
\end{lemma}

\bibliographystyle{IEEEtran}
\bibliography{string,references}

\end{document}